\documentclass[twoside,twocolumn,10pt]{article}
\usepackage{extsizes}
\usepackage[super,sort&compress,comma]{natbib} 
\usepackage[version=3]{mhchem}
\usepackage[left=1.5cm, right=1.5cm, top=1.785cm, bottom=2.0cm]{geometry}
\usepackage{balance}
\usepackage{widetext}
\usepackage{times,mathptmx}
\usepackage{sectsty}
\usepackage{graphicx} 
\usepackage{lastpage}
\usepackage[format=plain,justification=raggedright,singlelinecheck=false,font={stretch=1.125,small,sf},labelfont=bf,labelsep=space]{caption}
\usepackage{float}
\usepackage{fancyhdr}
\usepackage{fnpos}
\usepackage[english]{babel}
\usepackage{array}
\usepackage{droidsans}
\usepackage{charter}
\usepackage[T1]{fontenc}
\usepackage[usenames,dvipsnames]{xcolor}
\usepackage{setspace}
\usepackage[compact]{titlesec}
\usepackage{soul}

\definecolor{cream}{RGB}{222,217,201}

\begin{document}

\pagestyle{fancy}
\thispagestyle{plain}
\fancypagestyle{plain}{

%%%HEADER%%%
%\fancyhead[C]{\includegraphics[width=18.5cm]{1c.eps}}
%\fancyhead[L]{\hspace{0cm}\vspace{1.5cm}\includegraphics[height=30pt]{1a.eps}}
%\fancyhead[R]{\hspace{0cm}\vspace{1.7cm}\includegraphics[height=55pt]{1b.eps}}
\renewcommand{\headrulewidth}{0pt}
}
%%%END OF HEADER%%%

%%%PAGE SETUP - Please do not change any commands within this section%%%
\makeFNbottom
\makeatletter
\renewcommand\LARGE{\@setfontsize\LARGE{15pt}{17}}
\renewcommand\Large{\@setfontsize\Large{12pt}{14}}
\renewcommand\large{\@setfontsize\large{10pt}{12}}
\renewcommand\footnotesize{\@setfontsize\footnotesize{7pt}{10}}
\makeatother

\renewcommand{\thefootnote}{\fnsymbol{footnote}}
\renewcommand\footnoterule{\vspace*{1pt}% 
\color{cream}\hrule width 3.5in height 0.4pt \color{black}\vspace*{5pt}} 
\setcounter{secnumdepth}{5}

\makeatletter 
\renewcommand\@biblabel[1]{#1}            
\renewcommand\@makefntext[1]% 
{\noindent\makebox[0pt][r]{\@thefnmark\,}#1}
\makeatother 
\renewcommand{\figurename}{\small{Fig.}~}
\sectionfont{\sffamily\Large}
\subsectionfont{\normalsize}
\subsubsectionfont{\bf}
\setstretch{1.125} %In particular, please do not alter this line.
\setlength{\skip\footins}{0.8cm}
\setlength{\footnotesep}{0.25cm}
\setlength{\jot}{10pt}
\titlespacing*{\section}{0pt}{4pt}{4pt}
\titlespacing*{\subsection}{0pt}{15pt}{1pt}
%%%END OF PAGE SETUP%%%

%%%FOOTER%%%
\fancyfoot{}
%\fancyfoot[LO,RE]{\vspace{-7.1pt}\includegraphics[height=9pt]{2a.eps}}
%\fancyfoot[CO]{\vspace{-7.1pt}\hspace{13.2cm}\includegraphics{2a.eps}}
%\fancyfoot[CE]{\vspace{-7.2pt}\hspace{-14.2cm}\includegraphics{2a.eps}}
\fancyfoot[RO]{\footnotesize{\sffamily{1--\pageref{LastPage} ~\textbar  \hspace{2pt}\thepage}}}
\fancyfoot[LE]{\footnotesize{\sffamily{\thepage~\textbar\hspace{3.45cm} 1--\pageref{LastPage}}}}
\fancyhead{}
\renewcommand{\headrulewidth}{0pt} 
\renewcommand{\footrulewidth}{0pt}
\setlength{\arrayrulewidth}{1pt}
\setlength{\columnsep}{6.5mm}
\setlength\bibsep{1pt}
%%%END OF FOOTER%%%

%%%FIGURE SETUP - please do not change any commands within this section%%%
\makeatletter 
\newlength{\figrulesep} 
\setlength{\figrulesep}{0.5\textfloatsep} 

\newcommand{\topfigrule}{\vspace*{-1pt}% 
\noindent{\color{cream}\rule[-\figrulesep]{\columnwidth}{1.5pt}} }

\newcommand{\botfigrule}{\vspace*{-2pt}% 
\noindent{\color{cream}\rule[\figrulesep]{\columnwidth}{1.5pt}} }

\newcommand{\dblfigrule}{\vspace*{-1pt}% 
\noindent{\color{cream}\rule[-\figrulesep]{\linewidth}{1.5pt}} }

\makeatother
%%%END OF FIGURE SETUP%%%

%%%TITLE, AUTHORS AND ABSTRACT%%%

\twocolumn[
  \begin{@twocolumnfalse}
\sffamily
\begin{tabular}{m{0.01cm} p{17.99cm} }
& \noindent\LARGE{\textbf{Universal scaling in active single-file dynamics
}} \\
\vspace{0.3cm} & \vspace{0.3cm} \\
& \noindent\large{Pritha Dolai$^\ast$, Arghya Das, Anupam Kundu, Chandan Dasgupta, Abhishek Dhar and K. Vijay Kumar$^\dagger$} \\
\vspace{2.5cm}  & \vspace{0.3cm} \\
& \noindent\normalsize{We study the single-file dynamics of three classes of active particles: run-and-tumble particles, active Brownian particles and active Ornstein-Uhlenbeck particles. At high activity values, the particles, interacting via purely repulsive and short-ranged forces, aggregate into several motile and dynamical clusters of comparable size, and do not display bulk phase-segregation.  In this dynamical steady-state, we find that the cluster size distribution of these aggregates is a scaled function of the density and activity parameters across the three models of active particles with the same scaling function. The velocity distribution  of these motile clusters is non-Gaussian. We show that the effective dynamics of these clusters can explain the observed emergent scaling of the mean-squared displacement of tagged particles for all the three models with identical scaling exponents and functions. Concomitant with the clustering seen at high activities, we observe that the static density correlation function displays rich structures, including multiple peaks that are reminiscent of particle clustering induced by effective attractive interactions, while the dynamical variant shows non-diffusive scaling. Our study reveals a universal scaling behavior in the single-file dynamics of interacting active particles.} 
\end{tabular}
\end{@twocolumnfalse} \vspace{0.6cm}
]
%%%END OF TITLE, AUTHORS AND ABSTRACT%%%

%%%FONT SETUP - please do not change any commands within this section
\renewcommand*\rmdefault{bch}\normalfont\upshape
\rmfamily
\section*{}
\vspace{-1cm}

%%%FOOTNOTES%%%
\footnotetext{International Centre for Theoretical Sciences, Tata Institute of Fundamental Research, Hesaraghatta Hobli, Bengaluru North, India 560089}
\footnotetext{$^\ast$pritha.dolai@icts.res.in, $^\dagger$vijaykumar@icts.res.in}

%Please use \dag to cite the ESI in the main text of the article.
%If you article does not have ESI please remove the the \dag symbol from the title and the footnotetext below.
%\footnotetext{\dag~Electronic Supplementary Information (ESI) available: [details of any supplementary information available should be included here]. See DOI: 10.1039/b000000x/}
%additional addresses can be cited as above using the lower-case letters, c, d, e... If all authors are from the same address, no letter is required

%\footnotetext{\ddag~Additional footnotes to the title and authors can be included \emph{e.g.}\ `Present address:' or `These authors 
%contributed equally to this work' as above using the symbols: \ddag, \textsection, and \P. Please place the appropriate symbol next 
%to the author's name and include a \texttt{\textbackslash footnotetext} entry in the the correct place in the list.}

%%%END OF FOOTNOTES%%%

%%%MAIN TEXT%%%%

\section*{Introduction}
\label{section:intro}

Active matter is a novel class of driven nonequilibrium systems wherein the  consumption and dissipation of energy occurs at the level of the individual units
\cite{Bechinger,rmp,Sriram1}. This nonequilibrium driving breaks detailed balance and can lead to the emergence of surprising phenomena at large-scales.
Interacting systems of active particles are known to phase-separate \cite{Tailleur1,fily1,fily2}, show collective flocking states
\cite{Vicsek,Aitor,Nitin} and do not have a thermodynamic equation of state \cite{Solon,Solon2}. Although much work has been done at the continuum level, deciphering the nonequilibrium statistical mechanics of interacting self-propelled active particles remains a largely open problem. 
 
Scalar active matter generically consist of self-propelled particles that interact with other particles via isotropic interactions while their internal orientational degrees of freedom control the velocity of self-propulsion. In other words, the interactions between these 
scalar active particles do not depend on the orientations of the interacting pair of particles. Within this framework, three broad classes of scalar active particles have been considered \cite{Tailleur1,fily1,Tailleur2,fodor,Solon3,howse2007,palacci2010,Das2018,Kurzthaler2018,Malakar2020}: (1) run-and-tumble particles
(RTPs), (2) active Brownian particles (ABPs), and (3) active Ornstein-Uhlenbeck particles (AOUPs). These three classes are used as archetypal models in deciphering the nonequilibrium statistical physics of active particles. The essential difference between the three models lies in the characteristics of the active driving forces.

Recent studies on the dynamics of swimming droplets in confined geometries \cite{shashi}, RTPs on a lattice \cite{Soto1,Evans},  and escape probability of ABPs in an open channel single-file driven by an external force \cite{Orlandini} have demonstrated rich emergent behavior in low-dimensional systems of interacting active particles. However, the statistical physics underlying single-file motion of active particles has not been explored extensively. For passive equilibrium systems, it is well known that the asymptotic behavior of the mean-squared displacement (MSD) of tagged particles, confined to move in a single-file, is sub-diffusive and is a scaled function of the density and single-particle diffusivity \cite{Pincus,Kirone,Tridib,Chaitra,Wei,Lutz,Kollmann,Misko,Lizana2008}. What is the statistical physics of a single file of active particles? Are there scaling relations for the MSD of tagged active particles? It is well known that active particles cluster at high activity values in two dimensions \cite{Tailleur1}. How does this clustering relate to the MSD of tagged particles? Does clustering lead to non-trivial features in the (static and dynamic) density correlation functions? Are there universal statistical scaling behaviors among the three classes of interacting scalar active particles confined to move in one-dimension?

In this study, we consider the statistical physics of active particles in a single file geometry with periodic boundaries. We compare and contrast the dynamics of the three classes of active particles mentioned above in this geometry. Our active particles are free to move continuously in one-dimension while preserving their ordering. The main result of our study is the following: at high activity, the particles aggregate into \emph{dynamic motile clusters} of comparable sizes and we find that the cluster size distributions (CSD), the MSD of tagged particles, and density correlation functions display universal scaling, both with respect to scaling exponents and scaling functions, across the three models. We observe that the effective degrees of freedom in this single-file geometry are the motile clusters of active particles, and provide a heuristic analysis of the dynamics of these clusters that can rationalize the observed universal scaling.

\section*{A single-file of active particles}
\label{section:model}

\begin{figure*}
\includegraphics[width=\linewidth]{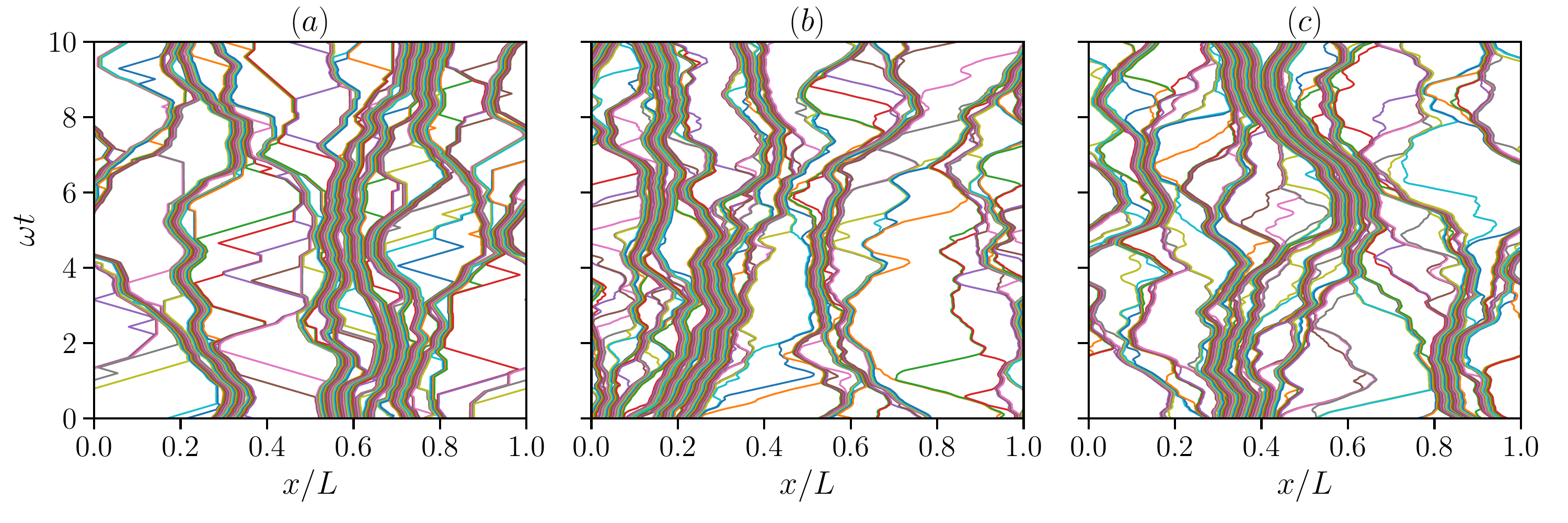}
\caption{Space-time trajectories of $N=100$ active particles, shown by different colors, in a one-dimensional periodic domain at a number-density $\rho=0.3$ (which leads to a domain size of $L=300$). Notice the similarity in the dynamics of motile clusters for the case of interacting (a) RTPs, (b) ABPs, and (c) AOUPs. The translational diffusion coefficient $D=0$, and the equivalent parameters, as defined in equation \eqref{equivalentParameters}, are $u=1$ and $\omega=10^{-2}$.} \label{trajectories}
\end{figure*}

We consider a system of $N$ interacting active particles confined to move on a one-dimensional line with periodic boundary conditions. The equation of motion for the $i^{\rm th}$ active particle, with position coordinate $x_i$ at time $t$, is
\begin{align}
\frac{d x_i}{dt} = v_i + \mu \, \left( F_{i,i+1} + F_{i,i-1} \right) 
+ \sqrt{2D} \, \eta_i(t)
\label{eom}
\end{align}
where $v_i$ is the active velocity, $\mu$ is the translational mobility, 
$D$ is the translational diffusion coefficient, and $\eta_i$ is a zero-mean unit-variance and uncorrelated Gaussian white noise process.
The $i^{\rm th}$ active particle interacts only with its nearest neighbours via the forces $F_{i, i\pm 1}$ derived from a purely repulsive Weeks-Chandler-Anderson 
(WCA) potential \cite{wca}
\begin{align}
U(r)=\epsilon \left\{ \begin{array}{lr}  \frac{1}{4}+ \left(\frac{\sigma}{r}\right)^{12}- \left(\frac{\sigma}{r}\right)^{6}, &  r<a
\\ 0, & r>a
\end{array}
\right.
\label{eq:wca}
\end{align}
where $\epsilon$ and $\sigma$ are the energy and length scales of the interaction respectively, and the cutoff distance $a=2^{1/6}\sigma$ sets the effective size of the particles.

The three classes of active particles are distinguished by the nature of their active velocity $v_i$. Each RTP has a bounded and discrete active velocity $v_i^{\rm RTP}$ that flips between
\begin{align}
v^{\rm RTP}_i = \pm v_R
\label{rtp}
\end{align}
at a Poisson rate $\gamma$ where $v_R$ is a fixed active speed. On the other hand, the active velocity $v_i^{\rm ABP}$ of each ABP is set by an internal angular coordinate $\theta_i$ that itself undergoes rotational diffusion. The evolution of $\theta_i$ and $v_i^{\rm ABP}$ are governed by
\begin{align}
v^{\rm ABP}_i = v_A \, \cos\theta_i,
\qquad
\frac{d\theta_i}{dt} = \sqrt{2D_r} \; \zeta_i(t)
\label{abp}
\end{align}
where $D_r$ is a rotational diffusion constant and $\zeta_i(t)$ is an uncorrelated Gaussian white noise process with zero-mean and unit-variance. Thus, ABPs have a bounded but continuous distribution of active velocities with a characteristic active speed $v_A$. The third class of active particles that we consider, namely AOUPs, have an active velocity $v_i^{\rm AOUP}$ that arises from an Ornstein-Uhlenbeck process satisfying
\begin{align}
\tau \frac{d v_i^{\rm AOUP}}{dt} = - v_i^{\rm AOUP} + \sqrt{2 \Delta} \; \xi_i(t)
\label{aoup}
\end{align}
where $\tau$ is a persistence-time constant, $\Delta$ is the strength of the noise-term and $\xi_i(t)$ is an uncorrelated Gaussian white noise process with zero-mean and unit-variance. Thus, the active velocity of an AOUP is a continuous but unbounded random variable. A characteristic active speed of AOUPs is given by $v_O =\sqrt{\Delta/\tau}$. Notice that, even though the active velocities, $v_i^{\rm RTP}$, $v_i^{\rm ABP}$ and $v_i^{\rm AOUP}$, are drawn from different distributions, their two-point time-correlation, for each model, has an exponential decay in all the three cases. This ensures that the long-time dynamics of all the above three classes of active-particles approach that of a passive Brownian particle in the limit of small persistence time, which corresponds to $\gamma \to \infty,~D_r \to \infty ~\mbox{and}~\tau \to 0$ for RTPs, ABPs and AOUPs respectively.

We first characterize the dynamics of an unconfined and non-interacting active particle in one dimension. The MSD $\langle \Delta x^2 \rangle$ for the three classes of active particles are \cite{Kanaya,Bechinger}
\begin{align}
\langle \Delta x^2 \rangle_{\rm RTP} &= \left(\frac{v_R}{\gamma}\right)^2 \, \left(\gamma \, t -\frac{1-e^{-2\gamma \, t}}{2} \right)+2D t, \label{freertp}
\\
\langle \Delta x^2 \rangle_{\rm ABP} &= \left( \frac{v_A}{D_r} \right)^2 \left(D_r t -1 + e^{-D_r t} \right)+2D t, \label{freeabp}
\\
\langle \Delta x^2 \rangle_{\rm AOUP} &= 2\,\tau \, \Delta\, \left(\frac{t}{\tau} - 1 + e^{-t/\tau} \right)+2D t,
\nonumber \\
&= 2(v_O \tau)^2 \left(\frac{t}{\tau} - 1 + e^{-t/\tau} \right)+2D t. \label{freeaoup}
\end{align}
where $\langle \cdots \rangle$ indicate noise averages. It is clear from the above equations that the MSD of free active particles is a scaled function of the corresponding scaled times ($\gamma \, t$, $D_r t$ and $t/\tau$) and persistence lengths ($v_R/\gamma$, $v_A/D_r$ and $v_O\tau$) for each class. In fact, introducing a rescaled speed $u$ and a persistence-rate $\omega$, and identifying
\begin{align}
v_R = u = v_O, \quad v_A = \sqrt{2} \, u,
\qquad
\gamma = \omega, \quad D_r = 2 \omega = \tau^{-1},
\label{equivalentParameters}
\end{align}
we notice that equations~\eqref{freertp}-\eqref{freeaoup} are scaled functions of $u$ and $\omega$ with the \emph{same scaling function}
\begin{align}
\langle \Delta x^2 \rangle = (u/\omega)^2 \, \left(\omega \, t -(1-e^{-2\omega \, t})/2 \right)+2D\,t.
\label{NI-AP-MSD-scaling}
\end{align}
Also notice that, for $D=0$, we have $\langle \Delta x^2 \rangle \sim (u\,t)^2$ at short times (compared to $\omega^{-1}$) and $\langle \Delta x^2 \rangle \sim (u^2/\omega) \, t$ at long times which allows us to identify the effective long-time diffusion constant as $u^2/(2\omega)$ \cite{Kanaya}. 

We note that the  universal scaling behavior in equation \eqref{NI-AP-MSD-scaling} is true for non-interacting particles in the absence of an external potential. In the presence of a confining potential it is known that the three classes can exhibit very different behaviour. For example it is known that RTPs and ABPs in harmonic traps can exhibit passive-to-active crossovers whereby  their steady-state density distributions change from being peaked at the centre to developing off-centre peaks \cite{Takatori2016,Dhar2019,Malakar2020} --- in contrast AOUPs are always described by centrally peaked Gaussian distributions \cite{Das2018,Szamel}. A natural  question, then, is whether a  universal scaling descriptions exists for the statistical properties of  active particles in the presence of interactions.

The main result of this study is that the cluster-size distribution, the mean-squared-displacements of tagged particles and the (static and dynamic) density-correlation functions of a single-file of $N$ interacting active particles in a periodic box of size $L$ are scaled functions of the number-density $\rho=N/L$ and equivalent activity parameters ($u$ and $\omega$) across the three models of active particles. Note that equation~\eqref{equivalentParameters} defines a correspondence to identify equivalent parameters across the three models, and we thus use $u$ and $\omega$ in the rest of the paper to present our scaling results.

To study the statistical dynamics of interacting active particles, we perform explicit Langevin simulations of the equation of motion \eqref{eom} and the evolution equation for the active velocity, equations \eqref{rtp}-\eqref{aoup}, for the corresponding model. We use $\sigma$, $\sigma^{2}/\mu \varepsilon$ and $\varepsilon/\sigma$ as the units of length, time and force respectively, employ an explicit Euler-Maruyama algorithm \cite{KloedenPlaten} to integrate the Langevin equations with a constant {time step $\Delta t = 10^{-4}$} and extract the various statistical distributions in the steady-state by averaging over {$10^3$ noise realizations}.

\subsection*{Particle trajectories and clusters}
\label{subsection:trajectoriesClusters}

In Fig.~\ref{trajectories}, we display the trajectories of {$N=100$} interacting active particles in a one-dimensional periodic box with a number density $\rho=N/L=0.3$ for the three classes (RTP, ABP and AOUP). The dimensionless equivalent parameters chosen were $u=1$ and $\omega=10^{-2}$, and the translational diffusion coefficient $D=0$. By a careful observation of the particle trajectories, we deduce the following points. 

First, the active particles form clusters. This is akin to the phenomenon of motility-induced-phase-separation (MIPS) seen in two-dimensional studies of interacting active particles at high activity \cite{Tailleur1,Bechinger}. However, unlike in two-dimensional systems, the aggregates of active particles seen in Fig.~\ref{trajectories} do not form a single big cluster. Rather, we see several clusters of similar size. Note that the active particles will form clusters only if they collide before their velocities change sign. In other words, clustering occurs only if the typical collision frequency $\rho u$ is larger than the typical rate $\omega$ at which the active velocities flip their directions, \emph{i.e.}, if $\rho u/\omega \gg 1$. On the other hand, if $\rho u/\omega \ll 1$, the particles reorient much faster than they collide with each other and thus do not form clusters. In this case, the particle dynamics will be very similar to those of a system of passive Brownian particles moving in a single-file. Thus, a large-persistence length of the active particles ($\sim u/\omega$) compared to inter-particle spacing ($\sim \rho^{-1}$) leads to particle clustering in our system.

\begin{figure*}[h]
\centering
\includegraphics[width=0.9\linewidth]{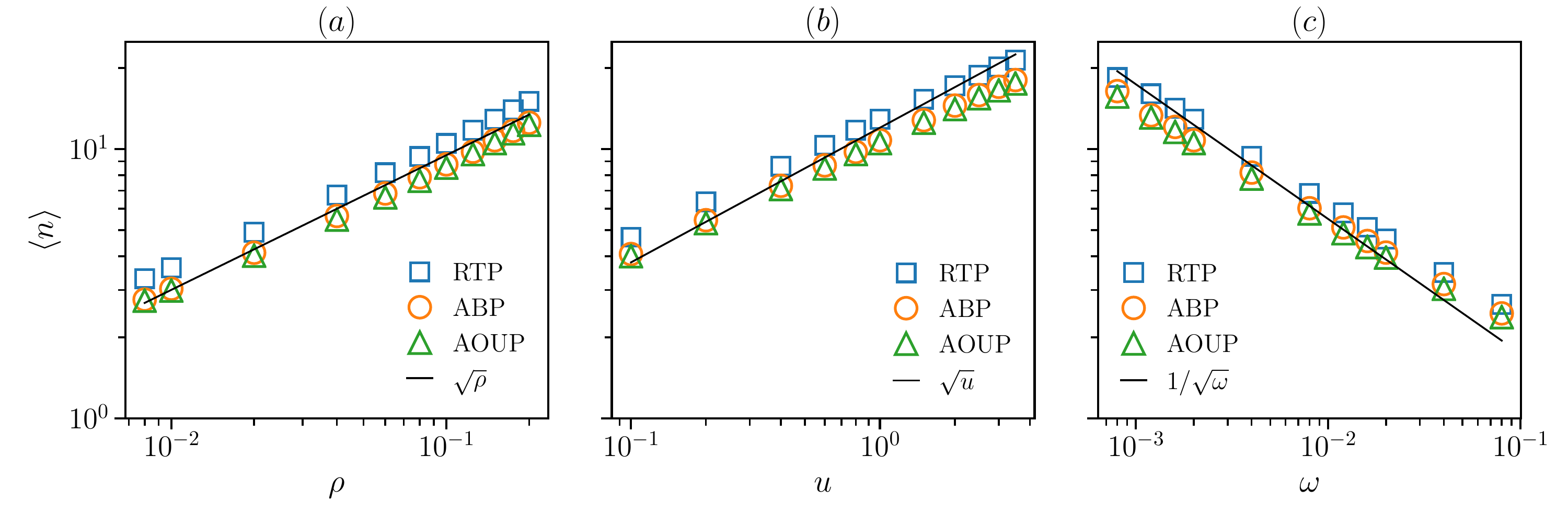}
\caption{The dependence of average cluster size $\langle n \rangle$ on (a) the density $\rho$, (b) the active speed $u$ and (c) the persistence-rate $\omega$ for RTPs, ABPs and AOUPs. In each plot, the unvaried parameters are fixed at $\omega=2 \times 10^{-3}$, $u=1$ and $\rho=0.15$. The solid lines are a guide to the eye. These plots suggest that the average cluster size $\langle n \rangle \sim \sqrt{\rho u/\omega}$.}\label{avgclsize}
\end{figure*}

Second, some of the clusters are motile. How do we understand this? From equation \eqref{eom}, the centre-of-mass $x_c$ ($=\sum_i^n x_i/n$) of a cluster of $n$ particles (with $D=0$) evolves according to $dx_c/dt = v_c$ where $v_c = \sum_{i=1}^n v_i/n$ is the centre-of-mass velocity of the cluster. Thus, if $v_c \neq 0$, the cluster is a motile object. The emergence of moving clusters in our continuum description is in sharp contrast to lattice models of active particles with exclusion interactions where particles can only form static clusters\cite{Evans}. It should be noted that $v_c$ is itself a random variable and, as such, a stable cluster with $n$ particles can change its direction of motion in time due to an internal rearrangement of the active velocities of the particles contained in it. Thus, at large times, these clusters of active particles themselves form emergent active particles albeit with slower dynamics.

Third, the clusters associate to form larger aggregates and break-up into smaller sized ones. When does a stable cluster split apart into two clusters due to an internal rearrangements of the active velocities? Two active particles, with the first particle positioned to the left of the second and having velocities $v_1$ and $v_2$ respectively,  can form a stable cluster for all configurations of $v_i$ except when $v_1 < v_2$. As the active velocities $v_i$ of the particles comprising a cluster evolve independently of each other, certain configurations of the $v_i$ can thus destabilize a cluster made of more than two particles. Consider a cluster comprised of $n$ particles, and labelled such that $i=1$ corresponds to the leftmost particle and $i=n$ corresponds to the rightmost particle. An update to the active velocities $v_i$ will split this cluster at the location following the $m^{\rm th}$-particle if (i) $v_m<v_c$ and $v_{m+1}>v_c$, and (ii) $\left(\sum_{i=1}^m v_i\right) < m \, v_c$ and $\left(\sum_{i=m+1}^n v_i\right) > (n-m) \, v_c$. In other words, the $n$-particle cluster will split into two clusters made of $m$ and $(n-m)$-particles if the active velocities $v_i$ of the particles switch sign at the location following the $m^{\rm th}$-particle, and if the cumulative sum of the active velocities to the left and right of this splitting point are negative and positive respectively, where both criteria are evaluated in the centre-of-mass frame of the cluster. Notice that this criteria is akin to the conditions by which a two particle cluster can break-up.

Fourth, the MSD of tagged particles must be intimately related to the dynamics of these motile clusters. Any tagged particle that is trapped inside a stable cluster is carried along with it as long as that cluster exists. As the individual active particles are persistent random walkers, at high activities, \emph{i.e.}, when $\rho u/\omega \gg 1$, it is highly unlikely to find isolated particles for extended durations. Thus, the statistical dynamics of a tagged particle will essentially be governed by the statistical dynamics of the clusters in which they are buried. As such, any non-trivial properties of clusters will be reflected in the dynamics of tagged particles and their density correlations as well.

In the following sections, we compute the steady-state distributions of the sizes and centre-of-mass velocities of these clusters. To connect this cluster statistics to tagged particle MSD, we then develop a heuristic theory that relates the emergent (quasi-particle) dynamics of clusters to the way in which the average cluster size scales with $\rho u/\omega$. We then compare the results of this heuristic scaling theory with the MSD computed from our Langevin simulations. Finally, we will discuss scaling behavior seen in two-point correlations of the fluctuating density field.

\section*{Statistics of clusters}
\label{section:clustering}

The trajectories of active particles shown in Fig.~\ref{trajectories} suggest a strong tendency of the active particles to cluster. As mentioned above, the particles form clusters for $\rho u/\omega \gg 1$. Our criteria for determining if two particles $i$ and $j$ form a cluster is whether $|x_i-x_j| < a$, \emph{i.e.}, if $i$ and $j$ are interacting via $U(r)$ and feel a repulsive force, we consider them to form a two-particle cluster in our numerics. Note that the statistical measures of clustering that we present below are computed in the steady-state by allowing the system to relax for long times $t \gg 1/\omega$ starting from a random initial condition.

\begin{figure*}[th]
\includegraphics[width=\linewidth]{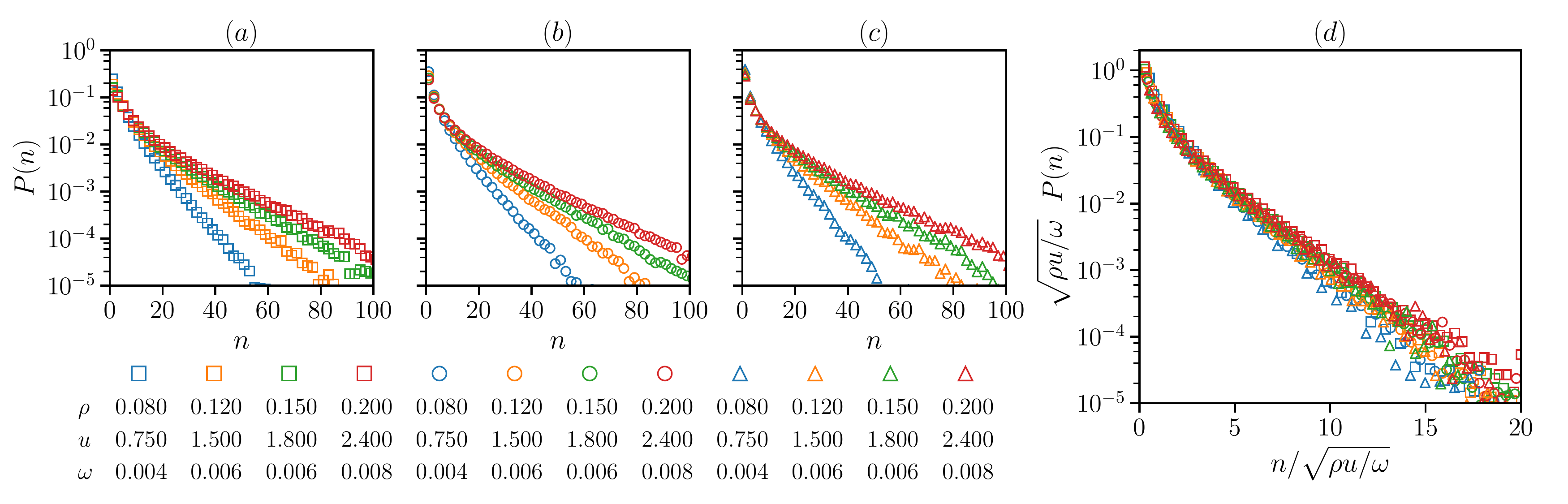}
\caption{Cluster size distribution of (a) RTPs, (b) ABPs and (c) AOUPs for a system with $N=450$ particles and various parameter values. We note that $P(n)$ is a monotonically decaying function of $n$ and has an exponential tail at large $n$ in each case. In (d), we re-plot the distributions in (a-c) by scaling them as per our conjecture \eqref{csdscaled} and find that the data show a good scaling collapse for a wide range of parameters.}
\label{csd}
\end{figure*}

With this criteria for clustering, we compute the average cluster size $\langle n \rangle$ as a function of the number-density $\rho$, the active speed $u$ and the velocity reorientation rate $\omega$.  The plots in Fig.~\ref{avgclsize} suggest that the average cluster size for all three classes of active particles behaves as
\begin{align}
\langle n \rangle \sim \sqrt{\frac{\rho u}{\omega}}
\label{aven}
\end{align}
in the regime where the active particles form clusters. Note that $\langle n \rangle \sim 1/\sqrt{\omega}$ is consistent with earlier studies \cite{Soto1}.

\begin{figure}[th]
\begin{center}
\includegraphics[width=\linewidth]{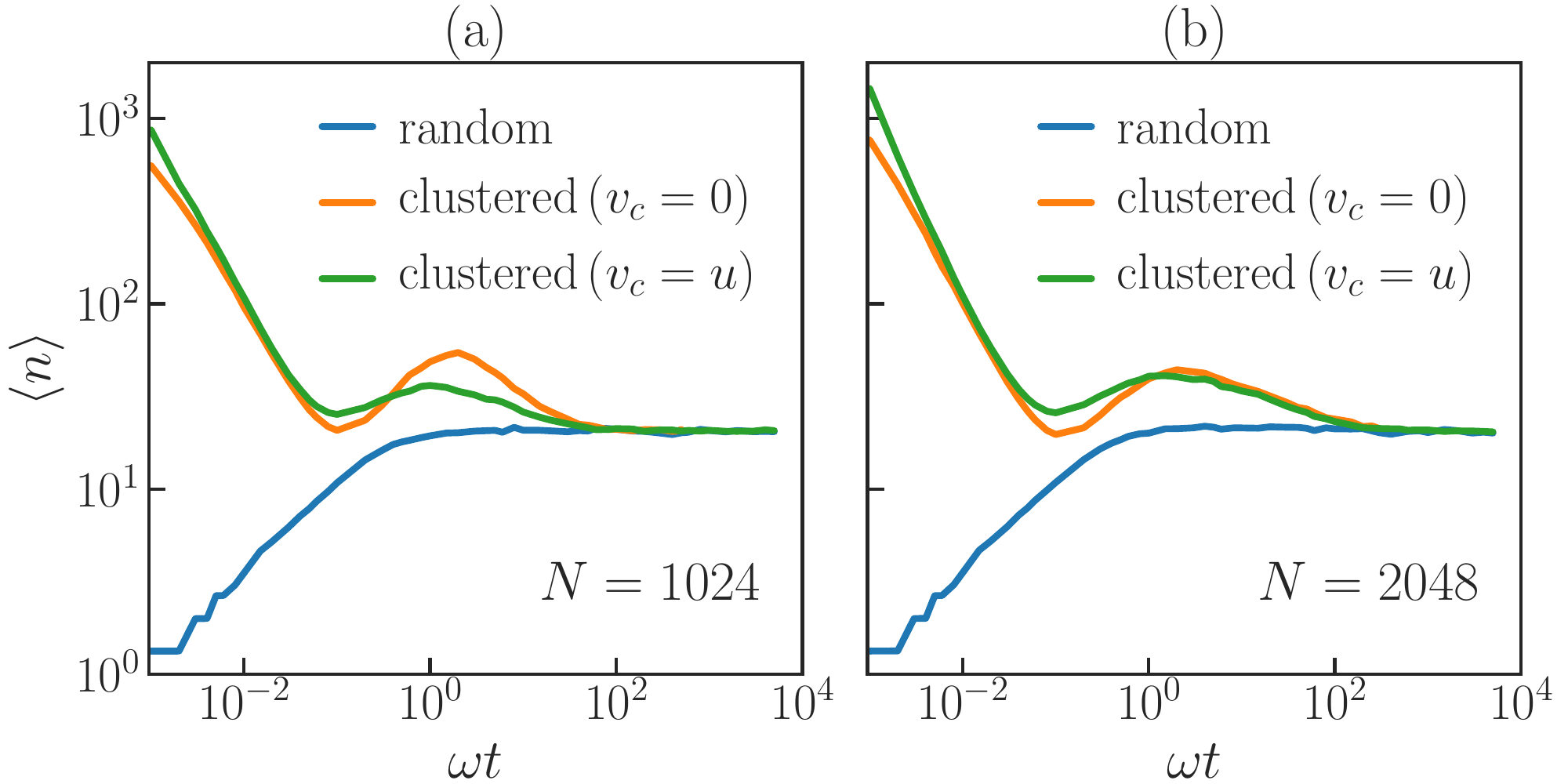}
\caption{Time-evolution of the average cluster size $\langle n \rangle$ for large system sizes $N$ of interacting RTPs (a) $N=1024$\,, (b) $N=2048$\,. We find that the asymptotic value $n^{\ast}$ does not scale with the system size and is also independent of the initial conditions. In particular, we find that the value of $n^{\ast}$ for initial conditions that start with all particles clustered is the same as that for initial conditions that start with random non-overlapping particle positions. Here we had $\rho=0.6$, $u=1.0$ and $\omega=0.005$.}
\label{aveN_vs_time}
\end{center}
\end{figure}

What is the distribution of cluster sizes across the three models and for various parameters? In Fig.~\ref{csd}(a-c), we show the CSD for the three classes of active particles. We note that (i) $P(n)$ for interacting active particles has an exponential decay at large $n$ \cite{fily2,Peruani}, and (ii) $P(n)$ is a monotonically decaying function of $n$ and does not peak at a large value of $n$, which is consistent with the fact that our one-dimensional system does not display any bulk phase-segregation. To confirm the non-existence of bulk phase-segregation in our single-file of active particles, we studied larger systems ($N=1024$ and $N=2048$) of RTPs in regimes of high activity $\rho u/\omega \gg 1$. We found that the asymptotic value of the average cluster size $\lim_{t \to \infty} \langle n \rangle = n^{\ast}$ approaches a finite value and does not scale with the system size. This asymptotic value $n^{\ast}$ is found to be independent of initial conditions, starting with either random non-overlapping particle positions or with an ordered cluster of particles, and depends only on the system parameters. We show the time-evolution of $\langle n \rangle$ in Fig.~\ref{aveN_vs_time}(a)-(b). Note that these results are consistent with previous studies \cite{Soto1,Evans} and also with coarse-grained theories that predict an initial coarsening stage without any true bulk-segregation asymptotically \cite{Tailleur1, cates_motility-induced_2015}. Rather, what is observed is a dynamical steady-state consisting of motile clusters that continuously aggregate and break-apart. We now characterize the statistical properties of cluster in this steady-state.

We now conjecture that the full steady-state CSD of active particles is a scaled function of $\sqrt{\rho u/\omega}$. With this conjecture, we found that the CSD for each model (RTP, ABP and AOUP) can be collapsed onto a single scaling function as shown in Fig.~\ref{csd}(d). Remarkably, we find that the CSD collapses onto the \emph{same scaling function} across all the three models that we have studied for several parameter values. In other words, our results suggest that the CSD for interacting active particles has the scaling form
\begin{align}
P(n) = \sqrt{\frac{\omega}{\rho u}} \;\; F\left(n \, \sqrt{\frac{\omega}{\rho u}}\right)
\label{csdscaled}
\end{align}
where $F$ is a universal scaling function. Note that since $\rho u / \omega$ is a dimensionless quantity, the pre-factor in the equation above is empirically determined by requiring a normalized $P(n)$. Equation \eqref{aven} suggests that we can also rewrite the above equation using the average cluster size $\langle n \rangle$ as $P(n)=1/\langle n \rangle \;  \mathcal{F}(n/\langle n \rangle)$. Note that our results in Fig~\ref{csd}(d) suggest that the scaling function $F$ is not a simple exponential. We would like to emphasize the remarkable scaling seen in Fig.~\ref{csd}(d) for all three classes of active particles, even though the individual particle dynamics for each model is driven by active (stochastic) processes with very different characteristics. In principle, each model of scalar active particles could have had a different scaling function. The fact that the data in Fig.~\ref{csd}(d) collapse onto a single master curve indicates a certain universality in the clustering statistics across the three models.

\begin{figure*}[h]
\includegraphics[width=\linewidth]{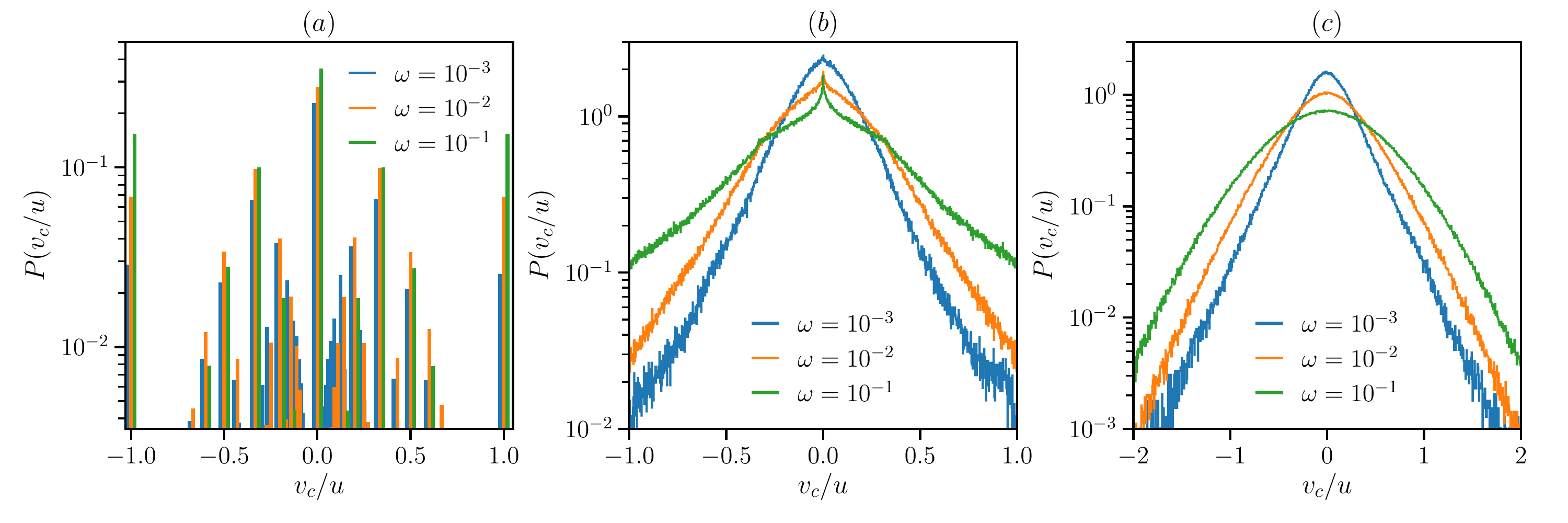}
\caption{(a) The cluster velocity distribution $P(v_c/u)$ for RTPs has a discrete spectrum of possible values for $v_c$ and shows dominant peaks at $v_c=0$ (corresponding to stationary clusters), $v_c=\pm u/3$ (corresponding to clusters with particle numbers $n=3,6,9,\ldots$) and $v_c=\pm u$ (corresponding to clusters in which all the RTPs have the same active velocity). Note that the plots for different $\omega$ are shifted for clarity. (b) For ABPs, the distribution $P(v_c/u)$ exhibits a cusp-like singularity at $v_c=0$ and a possible kink-like feature at $v_c/u=\pm1/3$. (c) For AOUPs, $P(v_c/u)$ is not Gaussian, even though the individual $v_i$ are Gaussian distributed. In all cases, $u=1$, $\rho=0.15$ and $N=150$. 
}
\label{cmv}
\end{figure*}

\subsection*{Cluster velocity distribution}
\label{subsection:clustervelocity}

The CSD shown in Fig.~\ref{csd} are at the steady-state and marginalized over the centre-of-mass velocity $v_c$ of the clusters. As is evident from Fig.~\ref{trajectories}, the clusters themselves display persistent motion. What is the distribution of their centre-of-mass velocities?

The active velocity of each RTP is one of the two discrete values $\pm u$. Consider a stable cluster of $n = n_+ + n_-$ particles, where $n_{\pm}$ are the number of particles in the cluster with velocities $\pm u$ respectively. The cluster velocity is then $v_c = u \, (n_{+} - n_{-})/n$. Thus the spectrum of $v_c$ will be discrete, $v_c/u \in [-1,1]$ with a step-size of $2/n$.
However, the statistical weight associated with each of these discrete values will be governed by the dynamics of clusters and is non-trivial to calculate. In Fig.~\ref{cmv}(a), we plot the probability distribution $P(v_c/u)$ of the centre-of-mass velocities of RTP clusters in the steady-state. Note that we have excluded isolated particles in constructing this distribution and also that $P(v_c/u)$ is marginalized over the size of the clusters. From Fig.~\ref{cmv}(a), we note that the distribution $P(v_c/u)$ is clearly different from a discrete Gaussian distribution. Instead, there are significant peaks at non-zero values of $v_c/u$. In particular, we notice dominant peaks at $v_c=0$, $v_c=\pm u$ and $v_c=\pm u/3$. The dominant peak at $v_c=0$ results from stationary stable clusters that have equal numbers of particles with oppositely directed active velocities. The peaks at $v_c=\pm u$ result from configurations in which all the particles constituting a cluster have the same active velocity. On the other hand, the significant peaks at $v_c=\pm u/3$ can only result from stable clusters containing $n=3,6,9,\ldots$ particles. We find that the primary contribution to these peaks arises from three-particle clusters. However, other clusters (with numbers in multiples of three) also contribute a non-zero weight to these peaks. We have checked that the qualitative features of $P(v_c/u)$ are independent of the number of particles $N$ in the system, even though the discrete spectrum has many more possible values.

The centre-of-mass velocity of an $n$-particle cluster of ABPs can take on a continuous set of values between $\pm u$. In Fig.~\ref{cmv}(b), we plot the cluster velocity distribution $P(v_c/u)$ for ABPs. Similar to the case of RTPs, we note that the distribution of $P(v_c/u)$ is non-Gaussian. However, there are no prominent peaks other than the central peak at $v_c=0$ corresponding to stationary clusters. Rather, we see a cusp-like singularity around $v_c=0$ and a kink-like feature around $v_c=\pm u/3$ in the distribution $P(v_c/u)$. Note that this kink-like feature is akin to prominent peaks seen in the case of RTPs at $v_c=\pm u/3$. Unlike the case of RTPs and ABPs, the velocity distribution for AOUP clusters is neither discrete nor bounded. Fig.~\ref{cmv}(c) shows the distribution $P(v_c/u)$ for AOUPs at different values of persistence-rates. The distribution, however, is not Gaussian, even though the active velocities of the individual AOUPs in the cluster are Gaussian random deviates. This is a result of the non-trivial constraints imposed by the stability criteria of a cluster.

In Fig. 4 (b), (c), we see that the $P(v_c/u)$ has a higher probability for larger values of $v_c/u$ as the flipping-rate $\omega$ increases. This can be understood as follows. The average cluster size $\langle n \rangle \sim \sqrt{\rho u /\omega}$ decreases with $\omega$ at fixed $\rho$ and $u$. As such, smaller sized clusters are more probable at larger $\omega$. For smaller sized clusters, it is more likely that their center-of-mass velocity will increase in magnitude. Conversely for small $\omega$, larger clusters are more likely and hence $P(v_c/u)$ will peak around zero. The probability of larger values of $v_c/u$, that require most particles in a cluster to have the same active speed, will then reduce.

The results presented above suggest that the statistical properties of clusters in the steady-state show non-trivial scaling properties. Importantly, the CSD collapses onto a single universal scaling function for various parameter values and also across the three models of active particles. Does this universal scaling extend to dynamical descriptors of the clustering of active particles?

\section*{Dynamics of tagged particles} 
\label{section:tagged}

The trajectories in Fig.~\ref{trajectories} provide an important insight to understand the MSD of tagged particles in our system. As remarked earlier, the dynamics of tagged particles is essentially governed by the dynamics of the clusters in which they are embedded. In other words, the clusters, on the timescales in which they do not break, can act as active `super-particles' that move ballistically at short time-scales (if $v_c\neq0$). On the same survival time-scales of the clusters, internal rearrangements of the active particles constituting the cluster can `flip' the direction of motion of the entire cluster at some effective rate $\Omega$. As such, for isolated clusters that survive for long times ($t \gg \Omega^{-1}$), we would expect to see emergent diffusive dynamics of the cluster as a whole.  This is similar to the dynamics of an active particle which reverses its persistent motion (with speed $v_c$) at a rate $\Omega$ and leads to an effective diffusion constant $v_c^2/\Omega$ at long-times \cite{Kanaya}. If, as argued above, the clusters as a whole behave like active super-particles, what is the effective persistence-rate $\Omega$ and the effective diffusion constant $D_{\rm eff}$ of a cluster of size $n$ moving with an active velocity $v_c$? Within an emergent picture of motile clusters, we now present a heuristic calculation for $\Omega$ and $D_{\rm eff}$, and relate it to tagged particle dynamics. For ease of this analysis, we consider the case of interacting RTPs. Similar considerations apply for the other two models. 

The effective velocity of a tagged particle embedded in an \emph{isolated} $n$-particle cluster of RTPs is just the centre-of-mass velocity of the cluster $v_c = \sum_{i=1}^n v_i/n$. Recall that each $v_i$ is an independent random variable that can flip between $v_i=\pm u$ at a rate $\omega$. In a coarse-grained description of this cluster as an effective RTP, what is the rate $\Omega$ at which $v_c$ flips sign? Notice that the range of $v_c/u$ is $[-1,1]$ with a step size of $2/n$. Each possible flip of the $v_i$ that does not break-apart the cluster can change $v_c$. Assuming, for the time-being, that an arbitrary configuration $\{v_i\}$ does not break-apart the $n$-particle cluster, the steady-state two-time correlation function of the centre-of-mass velocity is $\langle v_c(t)v_c(t') \rangle = (u^2/n) \; e^{-2\omega |t-t'|}$. At long times, this de-correlation of $v_c$ would lead to a diffusion of the centre-of-mass $x_c$ of the isolated cluster with an MSD $\langle \Delta x_c^2 (t)\rangle = 2 D_{\rm eff} \, t$ where the effective diffusion constant $D_{\rm eff} = u^2/(\omega \, n)$. Notice that this simple estimate for $D_{\rm eff}$ was arrived at by assuming that all possible velocity configurations are allowed, \emph{i.e.}, we included all internal rearrangements that could have either stabilized or destabilized the $n$-particle cluster. However, this estimate may not be accurate. To allow for a more realistic estimate, we make an ansatz that $D_{\rm eff} \sim v_c^2/\Omega \sim u^2/(\omega \, \langle n \rangle^{\alpha})$ where $\alpha$ is an unknown scaling exponent. We now outline a procedure to independently evaluate $D_{\rm eff}$, $v_c^2$ and $\Omega$ for the clusters, which will then allow us to infer the exponent $\alpha$. 

\begin{figure}[th]
\includegraphics[width=\linewidth]{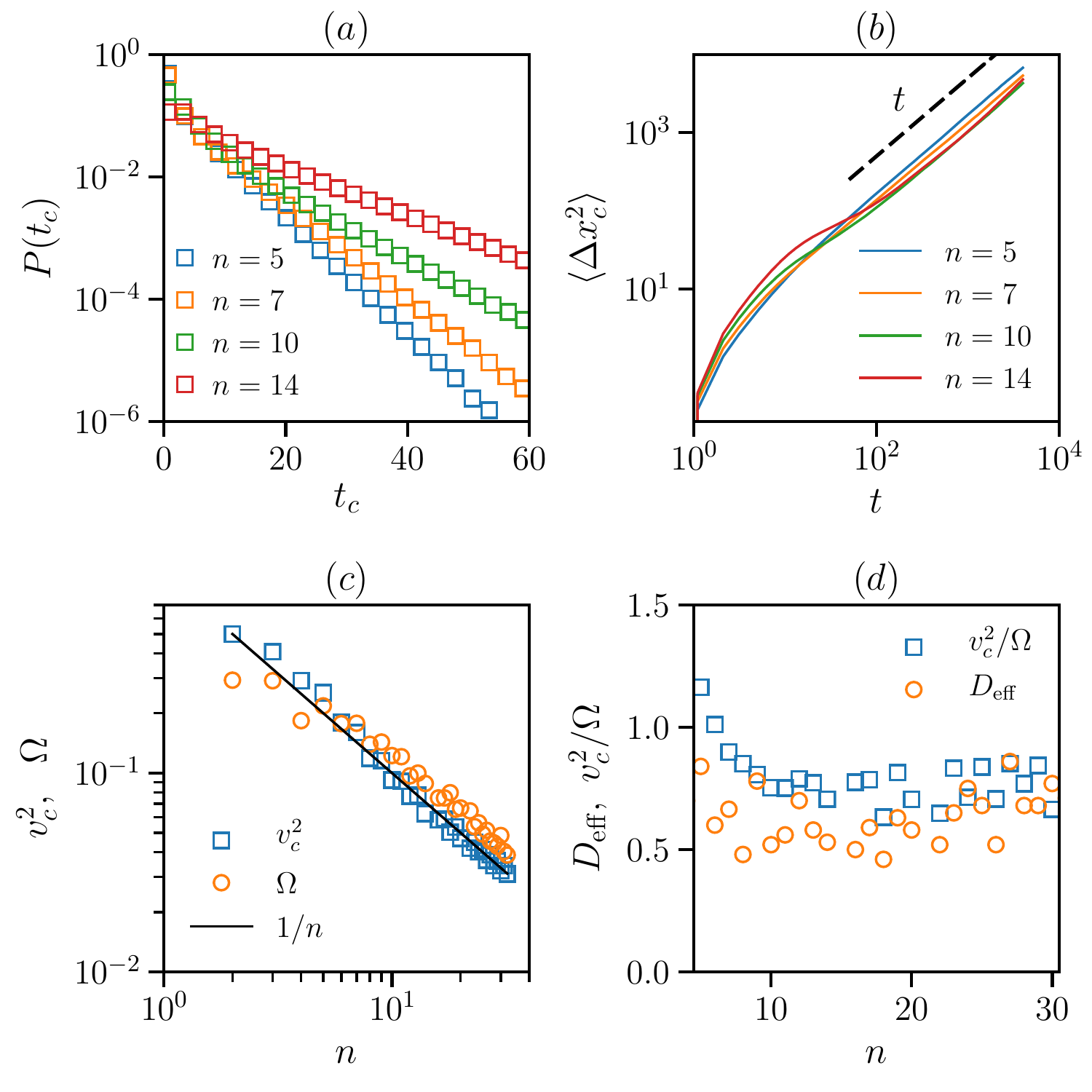}
\caption{Characterizing an isolated cluster of active particles using the dynamics outlined in the text. (a) For stable clusters, the tumble time distribution $P(t_c)$ of the cluster has an exponential tail. From this exponential tail, we extract the effective tumble-rate $\Omega$ of a cluster of size $n$. (b) The centre-of-mass MSD $\langle \Delta x_c^2 \rangle$ of an isolated cluster  of active particles shows a crossover from super-diffusive dynamics at short times to a diffusive behavior at long times. This asymptotic behavior of the MSD allows us to infer the effective diffusion coefficient $D_{\rm eff}$.
(c) The effective tumble-rate $\Omega$ and the square of center-of-mass velocity $v_c$ of an $n$-particle cluster scale as $\sim 1/n$. The solid line is a guide to the eye. (d) We compare $D_{\rm eff}$ inferred from (b) with $v_c^2/\Omega$ obtained from (c), and find that the effective diffusion coefficient $D_{\rm eff}$ is indeed proportional to $v_c^2/\Omega$ and is independent of the cluster size $n$, \emph{i.e.,} $\alpha \approx 0$.}
\label{Deff}
\end{figure}

Consider an isolated $n$-particle cluster of RTPs with each particle flipping its velocity between $v_i=\pm u$ at a Poisson rate $\omega$.  Starting from a stable configuration of velocities $\{v_i\}$, we time evolve this cluster by discarding updates to configurations that break it apart. After several such individual particle velocity flips, the cluster velocity can switch its direction amounting to a `tumble' of the cluster as a whole. Between two such successive tumbling events, the cluster moves persistently, albeit with an active speed that is no longer constant. In Fig.~\ref{Deff}(a), we plot the distribution $P(t_c)$ of tumble-times $t_c$ of an isolated RTP cluster of size $n$ as inferred from the dynamics outlined above. We find that $P(t_c)$ has an exponential tail at asymptotic values of $t_c$, \emph{i.e.}, $P(t_c) \sim \exp(-\Omega \, t_c)$ much like that of an individual RTP. This allows us to infer the tumble-rate $\Omega$ of this cluster. For times much longer than its tumble-rate $\Omega$, the centre-of-mass $x_c$ of an isolated cluster of active particles will diffuse with an effective diffusion constant $D_{\rm eff}$.  In Fig.~\ref{Deff}(b), we plot the MSD $\langle \Delta x_c^2 \rangle$ of isolated clusters of various sizes. We note that this MSD is super-diffusive at short times ($t \ll \Omega^{-1}$) and crosses over to a diffusive regime at long times. The asymptotic behavior of this MSD allows us to evaluate the effective diffusion constant $D_{\rm eff}$ of the cluster. Thus, from the above procedures, we can independently evaluate $v_c^2$, $\Omega$ and $D_{\rm eff}$ for an $n$-particle cluster of RTPs. In Fig.~\ref{Deff}(c), we plot the cluster size dependence of the centre-of-mass velocity and the effective flipping rate and find that $v_c^2 \sim 1/n$ and $\Omega \sim 1/n$. Finally, in  Fig.~\ref{Deff}(d),  we compare the measured values of $D_{\rm eff}$ and $v_c^2/\Omega$ for clusters of various sizes and find that they agree with each for a range of cluster sizes. We thus infer that the scaling exponent $\alpha \approx 0$. In other words, the effective diffusion constant of an isolated cluster of RTPs is 
\begin{align}
D_{\rm eff} \sim \frac{v_c^2}{\Omega} ~\approx ~{\rm constant},
\label{eq:Deff}
\end{align}
and is independent of the cluster size.

\begin{figure*}[ht]
\includegraphics[width=\linewidth]{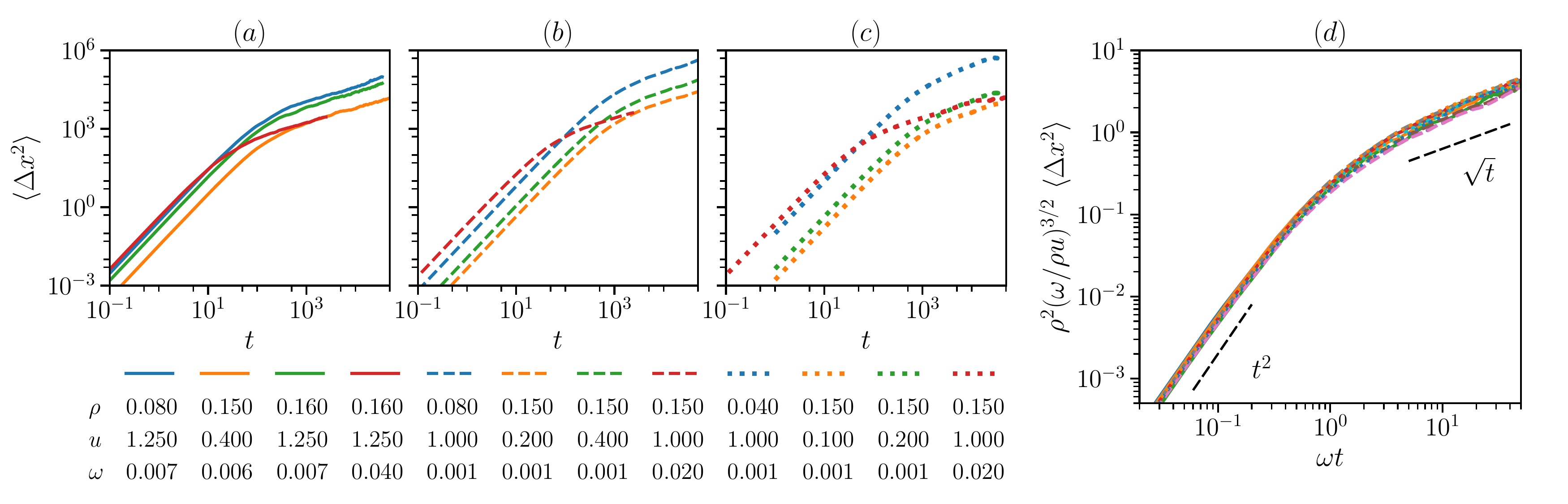}
\caption{The MSD of a tagged active particle $\langle \Delta x^2 \rangle$ calculated by numerically solving the Langevin equations \eqref{eom} for various parameter values in a system of $N=150$ particles in the case of (a) RTPs, (b) ABPs and (c) AOUPs. We notice that $\langle \Delta x^2 \rangle$ shows a crossover around $t \sim 1/\omega$ from a ballistic regime ($\sim t^2$) to a sub-diffusive regime ($\sim \sqrt{t}$) in all three cases. In (d), we rescale the time $t$ and $\langle \Delta x^2 \rangle$ as conjectured in \eqref{eq:msd} and find a remarkable scaling collapse, with the same scaling function, across all three models and for various parameter values.}
\label{msd}
\end{figure*}

We now return to the dynamics of tagged particles in our system. As discussed earlier, in a single file of active particles, the clusters act as active super-particles that can flip their motility directions and eventually diffuse on long time scales. From the preceding analysis, we have seen that the effective diffusion coefficient of an isolated cluster of active particles is independent of the cluster size.  Thus, the emergent picture for our single-file at long times is that of \emph{diffusing clusters moving in a single file}. It is well known\cite{Kirone,Tridib,Chaitra} that the MSD of a tagged 
particle in a single file of diffusing Brownian particles (with diffusion constant $D$ and particle-density $\rho$) has a scaling form  $\langle\Delta x^{2}\rangle = \sqrt{D \, t}/\rho$ in the sub-diffusive regime.  As such, for our diffusing clusters of active particles with effective diffusion constant $D_{\rm eff}$ and a cluster-density $\rho_{\rm cl} = \rho/\langle n \rangle$, we would expect that the MSD of tagged active particle clusters would be $\langle \Delta x^2\rangle 
\sim \sqrt{D_{\rm eff} \, t}/\rho_{\rm cl}$.
Using equations \eqref{aven} and \eqref{eq:Deff}, we thus get
\begin{align}
&\langle \Delta x^2 \rangle \sim  \sqrt{\frac{u^2 t}{\omega}} \, \frac{\langle n \rangle}{\rho}
\sim \sqrt{\frac{u^2 t}{\omega}} \, \frac{1}{\rho} \, \sqrt{\frac{\rho u}{\omega}} = \frac{1}{\rho^2} \left(\frac{\rho u}{\omega}\right)^{3/2} \, \sqrt{\omega t}. 
\label{MSD_asymp}
\end{align}
We now conjecture that Eq. \eqref{MSD_asymp} is the asymptotic limit of the following scaling form for the MSD of tagged active particles in a single file: 
\begin{align}
\langle\Delta x^{2}\rangle 
= \frac{1}{\rho^2} \, \left(\frac{\rho u}{\omega} \right)^{3/2} \,  G(\omega \, t),
\label{eq:msd}
\end{align}
where $G$ is a scaling function with an asymptotic behavior $G(x) \sim \sqrt{x}$ for $x\gg 1$. Notice that, by the above conjecture, we are anticipating that a single universal scaling function would govern the MSD of tagged particles across parameters and across the three models. How does this conjecture compare with the actual dynamics of active particles?

In Fig.~\ref{msd}(a-c), we plot the MSD $\langle \Delta x^2 \rangle$ of tagged particles for RTPs, ABPs and AOUPs as inferred from explicit numerical simulations of the Langevin equations \eqref{eom}. This MSD has a ballistic behavior ($\sim t^2$) at short times and crosses over to a sub-diffusive regime ($\sim \sqrt{t}$) at long times. This crossover from ballistic to sub-diffusive dynamics occurs around $t \sim \omega^{-1}$. Notice that we do not observe a diffusive regime ($\sim t$) in between. This is due to the fact that the parameters are such that typical collision rate between active particles exceeds the rate at which their velocities flip signs, \emph{i.e.}, $\rho u/\omega \gg 1$. This is the regime of parameters that leads to significant clustering.

We now plot the scaled MSD versus the scaled time in Fig.~\ref{msd}(d) and observe a remarkable scaling across parameters and the three models with the same universal scaling function $G$ as conjectured in equation \eqref{eq:msd}. What is even more remarkable is that there is a scaling collapse not just at long-times ($\omega t \gg 1$) as suggested by our analysis of cluster dynamics, but also at short times ($\omega t \ll 1$) wherein the clusters display super-diffusive motion. This also indicates that for $x \ll 1$, the scaling function $G(x) \sim x^2$. The observed collapse of tagged particle MSD, Fig.~\ref{msd}(d), across parameters and the three models strongly supports our picture of clusters acting as emergent particles in interacting active particle systems. It also supports our idea that a single dimensionless number, $\rho u / \omega$, essentially governs the dynamics of an active single-file. Note that, in addition to the data shown in Fig.~\ref{msd}, we have studied the tagged particle MSD for a much larger range of parameters and find the same scaling behavior.

\begin{figure*}[th]
\includegraphics[width=\linewidth]{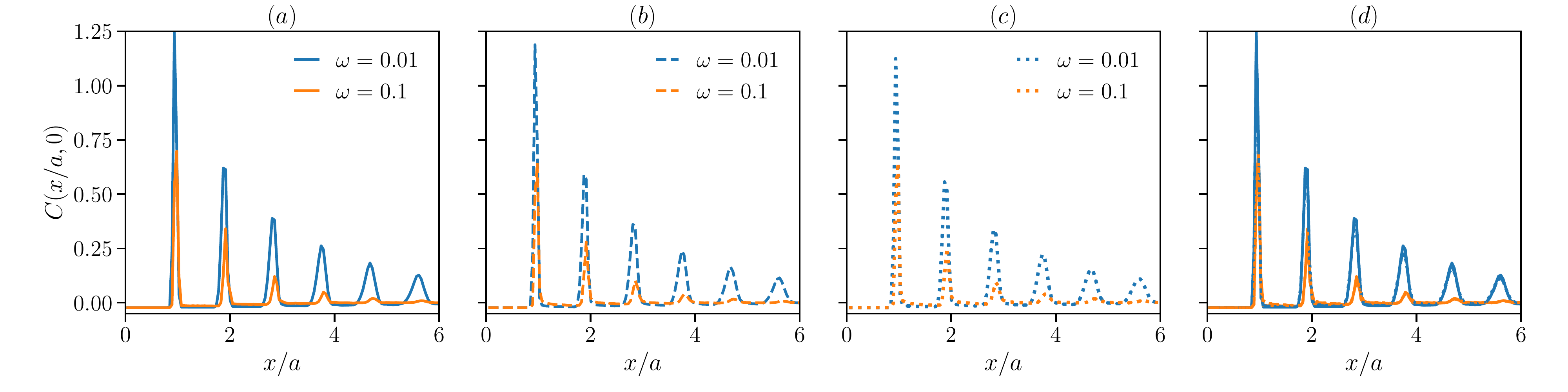}
\includegraphics[width=\linewidth]{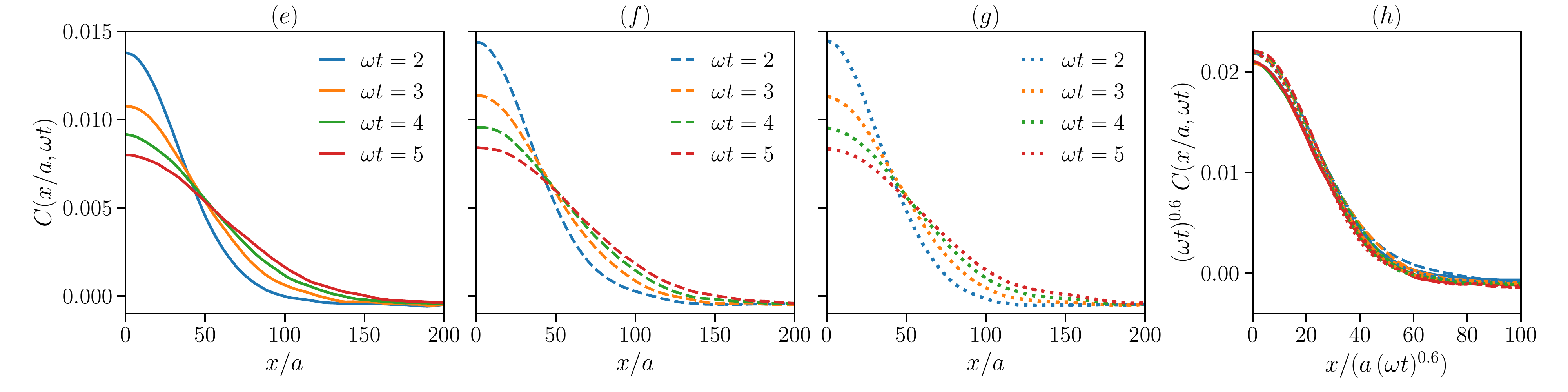}
\caption{
Two-point correlation function $C(x,t)$ of the fluctuating-density of active particles in RTPS (a,e), ABPs (b,f) and AOUPs (c,g). We see from (a-c) that the static density correlation $C(x,0)$ displays multiple peaks as a result of the clustering of active particles. We find that, for equivalent parameters, $C(x,0)$ for all three models lie on top of each other as seen in (d). In (a)-(d), we had $\rho=0.15$, $u=1.0$ and $N=150$. The dynamical density correlation $C(x,t)$ for RTPs (e), ABPs (f) and AOUPs (g) shows an approximate scaling collapse as indicated in equation \eqref{dynamic-density} and shown in (h). In (e)-(h), we had $\rho=0.15$, $u=1.0$ and $N=900$.}
\label{density-correlation}
\end{figure*}

\subsection*{Two-point correlations}

The interactions between the active particles in our simulations are governed by the purely repulsive WCA potential given in equation \eqref{eq:wca}. Yet, at high activity, the particles display clustering, indicating an effective attractive interaction between them. This is a well known fact in active particle systems which leads to the phenomenon of motility-induced phase separation \cite{Tailleur1,fily1,fily2}. A measure of these effective attractive interactions is the correlation functions of the fluctuating density field $\hat{\rho}(x,t) = \sum_{i=1}^N \delta(x-x_i(t))$.
We define $\delta \hat{\rho}(x,t) = \hat{\rho}(x,t) - \rho$ and compute the two-point correlation function $C(x,t) = \langle \delta \hat{\rho}(x,t)\delta \hat{\rho}(0,0) \rangle$ where $\rho=N/L$ is the mean particle-density in the system.

The static density-correlation, \emph{i.e.}, $C(x,0)$, displays rich structures, including several peaks resulting from particle clustering as seen in Fig.~\ref{density-correlation}(a-c). These  peaks are similar to those seen in passive diffusing systems with attractive interactions \cite{bishop1982collective,lepri2005one,lee2017cluster,Bagchi}. Notice that the range of this effective attractive interaction spans several particle sizes. In Fig.~\ref{density-correlation}(d), we show that the static density-correlation functions across the three different models collapse on top of each other for equivalent activity parameters.
In Fig.~\ref{density-correlation}(e-g), we plot  $C(x,t)$ for the three models of active particles and find that this two-point dynamical correlation function has an approximate scaling collapse, Fig.~\ref{density-correlation}(h), of the form
\begin{align}
C(x,t) = \frac{1}{(\omega \, t)^{\beta}} \; H\left( \frac{x}{a \; (\omega \, t)^{\beta}} \right)
\label{dynamic-density}
\end{align}
where $\beta \approx 0.6$ is a scaling exponent and $H$ is scaling function. Notice that the above scaling with time ($\sim t^{0.6}$) should be contrasted with the case of a single-file  passive interacting Brownian particles \cite{Pincus,lizana2009diffusion}. This non-diffusive scaling implies that the density correlations retain the signatures of activity even though the clusters exhibit diffusive motion at long times.

\section*{Discussion}

All the scaling results that we have presented are obtained by setting the translational diffusion constant $D=0$ and in the regime of high activity $\rho u / \omega \gg 1$. If $D=0$ but $\rho u / \omega \ll 1$, the active particles reorient before they encounter each other and this reduces the clustering significantly. The tagged-particle MSD would then crossover from a very short ballistic regime to a diffusive regime around $t \sim \omega^{-1}$, and then from this diffusive regime to a final sub-diffusive regime around $t\sim \omega/(\rho u)^2$. In fact, asymptotically we would expect to have $\langle \Delta x^2 \rangle \sim \frac{\sqrt{(u^2/\omega) \; t}}{\rho}$. The other case with $D \neq 0$, and $\rho u / \omega \ll 1$, would also reduce the clustering of particles. In this case, we expect that the tagged-particle MSD would display a diffusive to ballistic to diffusive and finally to a sub-diffusive crossover with an effective diffusion constant $D + u^2/(2\omega)$ at long-times. {For this case, we have checked that the MSD of tagged active particles does show universal scaling but only in the sub-diffusive regime.} Therefore, to underscore the effect of activity in the collective dynamics, we have chosen to mostly work in the `active regime' and have set the bare translational diffusion constant to zero.

The aggregation of active particles into clusters occurs when the velocity reorientation rate is small compared to the rate at which the particles collide with each other. We emphasize that, although large clusters appear in 1D, there is no clustering transition or bulk phase segregation, and in fact clusters of all sizes appear in the system as seen from the trajectories in Fig.~(\ref{trajectories}) and from the cluster size distributions in Fig.~(\ref{csd}). It is remarkable that a single dimensionless combination ($\rho u/\omega$) sets the average cluster size and also governs the scaling of the CSD across the three models with the same scaling function. Since our active particles are free to move in a one-dimensional continuous line, rather than being confined to lattice positions \cite{Evans}, they can form dynamic motile clusters. The motility of these clusters emerges from the intrinsic particle dynamics and is not artificially imposed\cite{campos}. However, the centre-of-mass velocity distributions of these clusters are different across the three models with distinct features for each model. For example, RTP clusters have a discrete set of possible values for $v_c$ and their distribution $P(v_c)$ has prominent peaks at nonzero values that would survive any simple coarse-grained description of the clusters. The scaling of the CSD observed in the three models and for various parameters (with the same scaling function) might be a distinctive feature of the single-file geometry rather than being something that could be observed across active particle models studied in higher dimensions \cite{Caprini}.

In the presence of excluded-volume interactions, the dynamical properties of interacting particles confined to one-dimension is very different from that in two or three dimensions. The exclusion interaction maintains the ordering of the particles in a single file. Consequently, for example, in a system of passive Brownian particles, the MSD of a tagged particle is sub-diffusive ($\sim \sqrt{t}$) at long times. This slowing down of particle dynamics in a single file should be contrasted with that in the two or three dimensions where the MSD of a tagged particle is always diffusive at long times. The present study concerned a similar situation for active particles. Our analysis of particle trajectories, such as those in Fig.\ref{trajectories}, lead us to conclude that the effective degrees of freedom in a single file of active particles are the clusters of active particles. As such, using a heuristic theory, we were able to link the effective flipping rate and the effective diffusion coefficient of an isolated cluster of active particles to the MSD of tagged particles in the single file. We found that this heuristic analysis is able to suggest an asymptotic sub-diffusive scaling ($\sim\sqrt{t}$) for the MSD of tagged active particles which is indeed confirmed by the results of our numerical simulations. In fact, the scaling ansatz was shown to work even for earlier times wherein the clusters (and the embedded particles) show super-diffusive dynamics.

Note that even though it looks like the one-dimensional ordering of particles is enough to give $\sim\sqrt{t}$ 
scaling for the MSD of tagged particles, the existence of large motile clusters marks a remarkable departure from passive diffusive systems. This departure is 
quantitatively observed in the $1/\sqrt{\rho}$ density dependence of tagged particle MSD for active particles in contrast to $1/\rho$ dependence for passive particles with similar repulsive interactions.

The activity induced effective attractive interactions manifests in the appearance of multiple-peaks in the static density-correlation function. For the case of dynamical correlations, we observe an approximate scaling collapse. Notice that this scaling form is distinctly different compared to  the case of passive particles which have a diffusive dynamical density-correlation function. It remains unclear if this approximate scaling implies that a single-file of active particles would belong to the Kardar-Parisi-Zhang class of nonequilibrium systems \cite{quastel2015one}.

In conclusion, we have demonstrated that a single file of active particles has universal scaling relations for the distribution of clusters sizes, the mean-square displacement of tagged particles and density correlations for various parameter values across the three models of RTPs, ABPs and AOUPs. These scaling relations could possibly be tested in experimental realizations of active particles, for instance in vibrated granular media. We believe that our study opens up new perspectives to understand scaling dynamics in systems of strongly interacting active particles.

\section*{Acknowledgements}
We thank S. Ramaswamy for fruitful discussions. P.D. thanks ICTS for financial support and computing facilities. A.D. acknowledges CEFIPRA postdoctoral fellowship hosted at ICTS-TIFR. A.K. would like to acknowledge support from the SERB Early Career Research Award ECR/2017/000634 from the Science and Engineering Research Board, Department of Science and Technology, Government of India. C.D. acknowledges support from the Department of Science and Technology, India. K.V.K.'s research is supported by the Department of Biotechnology, India, through a Ramalingaswami re-entry fellowship and by the Max Planck Society through a Max-Planck-Partner-Group at ICTS-TIFR. All simulations were performed on the ICTS clusters \emph{tetris}, \emph{mario}, \emph{shaarduula}, \emph{sharabha}, \emph{brenner} and \emph{pokemon}.  We acknowledge support of the Department of Atomic Energy, Government of India, under project no. 12-R\&D-TFR-5.10-1100.

\bibliography{asfd.bib}

\providecommand{\noopsort}[1]{}\providecommand{\singleletter}[1]{#1}%
\providecommand*{\mcitethebibliography}{\thebibliography}
\csname @ifundefined\endcsname{endmcitethebibliography}
{\let\endmcitethebibliography\endthebibliography}{}
\begin{mcitethebibliography}{48}
\providecommand*{\natexlab}[1]{#1}
\providecommand*{\mciteSetBstSublistMode}[1]{}
\providecommand*{\mciteSetBstMaxWidthForm}[2]{}
\providecommand*{\mciteBstWouldAddEndPuncttrue}
  {\def\EndOfBibitem{\unskip.}}
\providecommand*{\mciteBstWouldAddEndPunctfalse}
  {\let\EndOfBibitem\relax}
\providecommand*{\mciteSetBstMidEndSepPunct}[3]{}
\providecommand*{\mciteSetBstSublistLabelBeginEnd}[3]{}
\providecommand*{\EndOfBibitem}{}
\mciteSetBstSublistMode{f}
\mciteSetBstMaxWidthForm{subitem}
{(\emph{\alph{mcitesubitemcount}})}
\mciteSetBstSublistLabelBeginEnd{\mcitemaxwidthsubitemform\space}
{\relax}{\relax}

\bibitem[Bechinger \emph{et~al.}(2016)Bechinger, Leonardo, L{\"o}wen,
  Reichhardt, Volpe, and Volpe]{Bechinger}
C.~Bechinger, R.~D. Leonardo, H.~L{\"o}wen, C.~Reichhardt, G.~Volpe and
  G.~Volpe, \emph{Rev. Mod. Phys.}, 2016, \textbf{88}, 045006\relax
\mciteBstWouldAddEndPuncttrue
\mciteSetBstMidEndSepPunct{\mcitedefaultmidpunct}
{\mcitedefaultendpunct}{\mcitedefaultseppunct}\relax
\EndOfBibitem
\bibitem[Marchetti \emph{et~al.}(2013)Marchetti, Joanny, Ramaswamy, Liverpool,
  Prost, Rao, and Simha]{rmp}
M.~C. Marchetti, J.-F. Joanny, S.~Ramaswamy, T.~B. Liverpool, J.~Prost, M.~Rao
  and R.~A. Simha, \emph{Rev. Mod. Phys.}, 2013, \textbf{85}, 1143\relax
\mciteBstWouldAddEndPuncttrue
\mciteSetBstMidEndSepPunct{\mcitedefaultmidpunct}
{\mcitedefaultendpunct}{\mcitedefaultseppunct}\relax
\EndOfBibitem
\bibitem[Ramaswamy(2010)]{Sriram1}
S.~Ramaswamy, \emph{Annu. Rev. Condens. Matter Phys.}, 2010, \textbf{1},
  323\relax
\mciteBstWouldAddEndPuncttrue
\mciteSetBstMidEndSepPunct{\mcitedefaultmidpunct}
{\mcitedefaultendpunct}{\mcitedefaultseppunct}\relax
\EndOfBibitem
\bibitem[Tailleur and Cates(2008)]{Tailleur1}
J.~Tailleur and M.~E. Cates, \emph{Phys.\ Rev. Lett.}, 2008, \textbf{100},
  218103\relax
\mciteBstWouldAddEndPuncttrue
\mciteSetBstMidEndSepPunct{\mcitedefaultmidpunct}
{\mcitedefaultendpunct}{\mcitedefaultseppunct}\relax
\EndOfBibitem
\bibitem[Fily and Marchetti(2012)]{fily1}
Y.~Fily and M.~C. Marchetti, \emph{Phys.\ Rev. Lett.}, 2012, \textbf{108},
  235702\relax
\mciteBstWouldAddEndPuncttrue
\mciteSetBstMidEndSepPunct{\mcitedefaultmidpunct}
{\mcitedefaultendpunct}{\mcitedefaultseppunct}\relax
\EndOfBibitem
\bibitem[Fily \emph{et~al.}(2014)Fily, Henkes, and Marchetti]{fily2}
Y.~Fily, S.~Henkes and M.~C. Marchetti, \emph{Soft matter}, 2014, \textbf{10},
  2132\relax
\mciteBstWouldAddEndPuncttrue
\mciteSetBstMidEndSepPunct{\mcitedefaultmidpunct}
{\mcitedefaultendpunct}{\mcitedefaultseppunct}\relax
\EndOfBibitem
\bibitem[Vicsek \emph{et~al.}(1995)Vicsek, Czir{\'o}k, Ben-Jacob, Cohen, and
  Shochet]{Vicsek}
T.~Vicsek, A.~Czir{\'o}k, E.~Ben-Jacob, I.~Cohen and O.~Shochet, \emph{Phys.
  Rev. Lett.}, 1995, \textbf{75}, 1226\relax
\mciteBstWouldAddEndPuncttrue
\mciteSetBstMidEndSepPunct{\mcitedefaultmidpunct}
{\mcitedefaultendpunct}{\mcitedefaultseppunct}\relax
\EndOfBibitem
\bibitem[Martin-G{\'o}mez \emph{et~al.}(2018)Martin-G{\'o}mez, Levis,
  Diaz-Guileracd, and Pagonabarraga]{Aitor}
A.~Martin-G{\'o}mez, D.~Levis, A.~Diaz-Guileracd and I.~Pagonabarraga,
  \emph{Soft matter}, 2018, \textbf{14}, 2610\relax
\mciteBstWouldAddEndPuncttrue
\mciteSetBstMidEndSepPunct{\mcitedefaultmidpunct}
{\mcitedefaultendpunct}{\mcitedefaultseppunct}\relax
\EndOfBibitem
\bibitem[Kumar \emph{et~al.}(2014)Kumar, Soni, Ramaswamy, and Sood]{Nitin}
N.~Kumar, H.~Soni, S.~Ramaswamy and A.~K. Sood, \emph{N. Comm.}, 2014,
  \textbf{5}, 4688\relax
\mciteBstWouldAddEndPuncttrue
\mciteSetBstMidEndSepPunct{\mcitedefaultmidpunct}
{\mcitedefaultendpunct}{\mcitedefaultseppunct}\relax
\EndOfBibitem
\bibitem[Solon \emph{et~al.}(2015)Solon, Stenhammar, Wittkowski, Kardar, Kafri,
  Cates, and Tailleur]{Solon}
A.~P. Solon, J.~Stenhammar, R.~Wittkowski, M.~Kardar, Y.~Kafri, M.~E. Cates and
  J.~Tailleur, \emph{Phys. Rev. Lett.}, 2015, \textbf{114}, 198301\relax
\mciteBstWouldAddEndPuncttrue
\mciteSetBstMidEndSepPunct{\mcitedefaultmidpunct}
{\mcitedefaultendpunct}{\mcitedefaultseppunct}\relax
\EndOfBibitem
\bibitem[Solon \emph{et~al.}(2015)Solon, Fily, Baskaran, Cates, Kafri, Kardar,
  and Tailleur]{Solon2}
A.~P. Solon, Y.~Fily, A.~Baskaran, M.~E. Cates, Y.~Kafri, M.~Kardar and
  J.~Tailleur, \emph{Nature Phys.}, 2015, \textbf{11}, 673\relax
\mciteBstWouldAddEndPuncttrue
\mciteSetBstMidEndSepPunct{\mcitedefaultmidpunct}
{\mcitedefaultendpunct}{\mcitedefaultseppunct}\relax
\EndOfBibitem
\bibitem[Cates and Tailleur(2013)]{Tailleur2}
M.~E. Cates and J.~Tailleur, \emph{Europhys. Lett.}, 2013, \textbf{101},
  20010\relax
\mciteBstWouldAddEndPuncttrue
\mciteSetBstMidEndSepPunct{\mcitedefaultmidpunct}
{\mcitedefaultendpunct}{\mcitedefaultseppunct}\relax
\EndOfBibitem
\bibitem[Fodor \emph{et~al.}(2016)Fodor, Nardini, Cates, Tailleur, Visco, and
  van Wijland]{fodor}
E.~Fodor, C.~Nardini, M.~E. Cates, J.~Tailleur, P.~Visco and F.~van Wijland,
  \emph{Phys. Rev. Lett.}, 2016, \textbf{117}, 038103\relax
\mciteBstWouldAddEndPuncttrue
\mciteSetBstMidEndSepPunct{\mcitedefaultmidpunct}
{\mcitedefaultendpunct}{\mcitedefaultseppunct}\relax
\EndOfBibitem
\bibitem[Solon \emph{et~al.}(2015)Solon, Cates, and Tailleur]{Solon3}
A.~P. Solon, M.~E. Cates and J.~Tailleur, \emph{Eur. Phys. J. Special Topics},
  2015, \textbf{224}, 1231\relax
\mciteBstWouldAddEndPuncttrue
\mciteSetBstMidEndSepPunct{\mcitedefaultmidpunct}
{\mcitedefaultendpunct}{\mcitedefaultseppunct}\relax
\EndOfBibitem
\bibitem[Howse \emph{et~al.}(2007)Howse, Jones, Ryan, Gough, Vafabakhsh, and
  Golestanian]{howse2007}
J.~R. Howse, R.~A. Jones, A.~J. Ryan, T.~Gough, R.~Vafabakhsh and
  R.~Golestanian, \emph{Phys. Rev. Lett.}, 2007, \textbf{99}, 048102\relax
\mciteBstWouldAddEndPuncttrue
\mciteSetBstMidEndSepPunct{\mcitedefaultmidpunct}
{\mcitedefaultendpunct}{\mcitedefaultseppunct}\relax
\EndOfBibitem
\bibitem[Palacci \emph{et~al.}(2010)Palacci, Cottin-Bizonne, Ybert, and
  Bocquet]{palacci2010}
J.~Palacci, C.~Cottin-Bizonne, C.~Ybert and L.~Bocquet, \emph{Phys. Rev.
  Lett.}, 2010, \textbf{105}, 088304\relax
\mciteBstWouldAddEndPuncttrue
\mciteSetBstMidEndSepPunct{\mcitedefaultmidpunct}
{\mcitedefaultendpunct}{\mcitedefaultseppunct}\relax
\EndOfBibitem
\bibitem[Das \emph{et~al.}(2018)Das, Gompper, and Winkler]{Das2018}
S.~Das, G.~Gompper and R.~G. Winkler, \emph{New J. Phys.}, 2018, \textbf{20},
  015001\relax
\mciteBstWouldAddEndPuncttrue
\mciteSetBstMidEndSepPunct{\mcitedefaultmidpunct}
{\mcitedefaultendpunct}{\mcitedefaultseppunct}\relax
\EndOfBibitem
\bibitem[Kurzthaler \emph{et~al.}(2018)Kurzthaler, Devailly, Arlt, Franosch,
  Poon, Martinez, and Brown]{Kurzthaler2018}
C.~Kurzthaler, C.~Devailly, J.~Arlt, T.~Franosch, W.~C. Poon, V.~A. Martinez
  and A.~T. Brown, \emph{Phys. Rev. Lett.}, 2018, \textbf{121}, 078001\relax
\mciteBstWouldAddEndPuncttrue
\mciteSetBstMidEndSepPunct{\mcitedefaultmidpunct}
{\mcitedefaultendpunct}{\mcitedefaultseppunct}\relax
\EndOfBibitem
\bibitem[Malakar \emph{et~al.}(2020)Malakar, Das, Kundu, Kumar, and
  Dhar]{Malakar2020}
K.~Malakar, A.~Das, A.~Kundu, K.~V. Kumar and A.~Dhar, \emph{Phys. Rev. E},
  2020, \textbf{101}, 022610\relax
\mciteBstWouldAddEndPuncttrue
\mciteSetBstMidEndSepPunct{\mcitedefaultmidpunct}
{\mcitedefaultendpunct}{\mcitedefaultseppunct}\relax
\EndOfBibitem
\bibitem[Thutupalli \emph{et~al.}(2018)Thutupalli, Geyer, Singh, Adhikari, and
  Stone]{shashi}
S.~Thutupalli, D.~Geyer, R.~Singh, R.~Adhikari and H.~A. Stone, \emph{PNAS},
  2018, \textbf{112}, 5403\relax
\mciteBstWouldAddEndPuncttrue
\mciteSetBstMidEndSepPunct{\mcitedefaultmidpunct}
{\mcitedefaultendpunct}{\mcitedefaultseppunct}\relax
\EndOfBibitem
\bibitem[Soto and Golestanian(2014)]{Soto1}
R.~Soto and R.~Golestanian, \emph{Phys.\ Rev. E.}, 2014, \textbf{89},
  012706\relax
\mciteBstWouldAddEndPuncttrue
\mciteSetBstMidEndSepPunct{\mcitedefaultmidpunct}
{\mcitedefaultendpunct}{\mcitedefaultseppunct}\relax
\EndOfBibitem
\bibitem[Slowman \emph{et~al.}(2016)Slowman, Evans, and Blythe]{Evans}
A.~B. Slowman, M.~R. Evans and R.~A. Blythe, \emph{Phys.\ Rev. Lett.}, 2016,
  \textbf{116}, 218101\relax
\mciteBstWouldAddEndPuncttrue
\mciteSetBstMidEndSepPunct{\mcitedefaultmidpunct}
{\mcitedefaultendpunct}{\mcitedefaultseppunct}\relax
\EndOfBibitem
\bibitem[Locatelli \emph{et~al.}(2015)Locatelli, Baldovin, and
  Orlandini]{Orlandini}
E.~Locatelli, F.~Baldovin and E.~Orlandini, \emph{Phys.\ Rev. E.}, 2015,
  \textbf{91}, 022109\relax
\mciteBstWouldAddEndPuncttrue
\mciteSetBstMidEndSepPunct{\mcitedefaultmidpunct}
{\mcitedefaultendpunct}{\mcitedefaultseppunct}\relax
\EndOfBibitem
\bibitem[Alexander and Pincus(1978)]{Pincus}
S.~Alexander and P.~Pincus, \emph{Phys.\ Rev. B.}, 1978, \textbf{18},
  2011\relax
\mciteBstWouldAddEndPuncttrue
\mciteSetBstMidEndSepPunct{\mcitedefaultmidpunct}
{\mcitedefaultendpunct}{\mcitedefaultseppunct}\relax
\EndOfBibitem
\bibitem[Krapivsky \emph{et~al.}(2015)Krapivsky, Mallick, and Sadhu]{Kirone}
P.~L. Krapivsky, K.~Mallick and T.~Sadhu, \emph{J. Stat. Phys.}, 2015,
  \textbf{160}, 885\relax
\mciteBstWouldAddEndPuncttrue
\mciteSetBstMidEndSepPunct{\mcitedefaultmidpunct}
{\mcitedefaultendpunct}{\mcitedefaultseppunct}\relax
\EndOfBibitem
\bibitem[Krapivsky \emph{et~al.}(2014)Krapivsky, Mallick, and Sadhu]{Tridib}
P.~L. Krapivsky, K.~Mallick and T.~Sadhu, \emph{Phys.\ Rev. Lett.}, 2014,
  \textbf{113}, 078101\relax
\mciteBstWouldAddEndPuncttrue
\mciteSetBstMidEndSepPunct{\mcitedefaultmidpunct}
{\mcitedefaultendpunct}{\mcitedefaultseppunct}\relax
\EndOfBibitem
\bibitem[Hegde \emph{et~al.}(2014)Hegde, Sabhapandit, and Dhar]{Chaitra}
C.~Hegde, S.~Sabhapandit and A.~Dhar, \emph{Phys.\ Rev. Lett.}, 2014,
  \textbf{113}, 120601\relax
\mciteBstWouldAddEndPuncttrue
\mciteSetBstMidEndSepPunct{\mcitedefaultmidpunct}
{\mcitedefaultendpunct}{\mcitedefaultseppunct}\relax
\EndOfBibitem
\bibitem[Wei \emph{et~al.}(2000)Wei, Bechinger, and Leiderer]{Wei}
Q.-H. Wei, C.~Bechinger and P.~Leiderer, \emph{Science}, 2000, \textbf{287},
  625\relax
\mciteBstWouldAddEndPuncttrue
\mciteSetBstMidEndSepPunct{\mcitedefaultmidpunct}
{\mcitedefaultendpunct}{\mcitedefaultseppunct}\relax
\EndOfBibitem
\bibitem[Lutz \emph{et~al.}(2004)Lutz, Kollmann, and Bechinger]{Lutz}
C.~Lutz, M.~Kollmann and C.~Bechinger, \emph{Phys.\ Rev. Lett.}, 2004,
  \textbf{93}, 026001\relax
\mciteBstWouldAddEndPuncttrue
\mciteSetBstMidEndSepPunct{\mcitedefaultmidpunct}
{\mcitedefaultendpunct}{\mcitedefaultseppunct}\relax
\EndOfBibitem
\bibitem[Kollmann(2003)]{Kollmann}
M.~Kollmann, \emph{Phys.\ Rev. Lett.}, 2003, \textbf{90}, 180602\relax
\mciteBstWouldAddEndPuncttrue
\mciteSetBstMidEndSepPunct{\mcitedefaultmidpunct}
{\mcitedefaultendpunct}{\mcitedefaultseppunct}\relax
\EndOfBibitem
\bibitem[Nelissen \emph{et~al.}(2007)Nelissen, Misko, and Peeters]{Misko}
K.~Nelissen, V.~R. Misko and F.~M. Peeters, \emph{Europhys. Lett.}, 2007,
  \textbf{80}, 56004\relax
\mciteBstWouldAddEndPuncttrue
\mciteSetBstMidEndSepPunct{\mcitedefaultmidpunct}
{\mcitedefaultendpunct}{\mcitedefaultseppunct}\relax
\EndOfBibitem
\bibitem[Lizana and Ambj\"ornsson(2008)]{Lizana2008}
L.~Lizana and T.~Ambj\"ornsson, \emph{Phys. Rev. Lett.}, 2008, \textbf{100},
  200601\relax
\mciteBstWouldAddEndPuncttrue
\mciteSetBstMidEndSepPunct{\mcitedefaultmidpunct}
{\mcitedefaultendpunct}{\mcitedefaultseppunct}\relax
\EndOfBibitem
\bibitem[J.D.Weeks \emph{et~al.}(1971)J.D.Weeks, D.Chandler, and
  H.C.Andersen]{wca}
J.D.Weeks, D.Chandler and H.C.Andersen, \emph{J. Chem. Phys.}, 1971,
  \textbf{54}, 5237\relax
\mciteBstWouldAddEndPuncttrue
\mciteSetBstMidEndSepPunct{\mcitedefaultmidpunct}
{\mcitedefaultendpunct}{\mcitedefaultseppunct}\relax
\EndOfBibitem
\bibitem[Malakar \emph{et~al.}(2018)Malakar, Jemseena, Kundu, Kumar,
  Sabhapandit, Majumdar, Redner, and Dhar]{Kanaya}
K.~Malakar, V.~Jemseena, A.~Kundu, K.~V. Kumar, S.~Sabhapandit, S.~N. Majumdar,
  S.~Redner and A.~Dhar, \emph{J. Stat. Mech: Theory and Expt}, 2018,
  043215\relax
\mciteBstWouldAddEndPuncttrue
\mciteSetBstMidEndSepPunct{\mcitedefaultmidpunct}
{\mcitedefaultendpunct}{\mcitedefaultseppunct}\relax
\EndOfBibitem
\bibitem[Takatori \emph{et~al.}(2016)Takatori, {De Dier}, Vermant, and
  Brady]{Takatori2016}
S.~C. Takatori, R.~{De Dier}, J.~Vermant and J.~F. Brady, \emph{Nat. Commun.},
  2016, \textbf{7}, 10694\relax
\mciteBstWouldAddEndPuncttrue
\mciteSetBstMidEndSepPunct{\mcitedefaultmidpunct}
{\mcitedefaultendpunct}{\mcitedefaultseppunct}\relax
\EndOfBibitem
\bibitem[Dhar \emph{et~al.}(2019)Dhar, Kundu, Majumdar, Sabhapandit, and
  Schehr]{Dhar2019}
A.~Dhar, A.~Kundu, S.~N. Majumdar, S.~Sabhapandit and G.~Schehr, \emph{Phys.
  Rev. E}, 2019, \textbf{99}, 032132\relax
\mciteBstWouldAddEndPuncttrue
\mciteSetBstMidEndSepPunct{\mcitedefaultmidpunct}
{\mcitedefaultendpunct}{\mcitedefaultseppunct}\relax
\EndOfBibitem
\bibitem[Szamel(2014)]{Szamel}
G.~Szamel, \emph{Phys. Rev. E}, 2014, \textbf{90}, 012111\relax
\mciteBstWouldAddEndPuncttrue
\mciteSetBstMidEndSepPunct{\mcitedefaultmidpunct}
{\mcitedefaultendpunct}{\mcitedefaultseppunct}\relax
\EndOfBibitem
\bibitem[Kloeden()]{KloedenPlaten}
P.~E. Kloeden, \emph{Numerical Solution of Stochastic Differential
  Equations}\relax
\mciteBstWouldAddEndPuncttrue
\mciteSetBstMidEndSepPunct{\mcitedefaultmidpunct}
{\mcitedefaultendpunct}{\mcitedefaultseppunct}\relax
\EndOfBibitem
\bibitem[Peruani and B{\"a}r(2013)]{Peruani}
F.~Peruani and M.~B{\"a}r, \emph{New J. Phys.}, 2013, \textbf{15}, 065009\relax
\mciteBstWouldAddEndPuncttrue
\mciteSetBstMidEndSepPunct{\mcitedefaultmidpunct}
{\mcitedefaultendpunct}{\mcitedefaultseppunct}\relax
\EndOfBibitem
\bibitem[Cates and Tailleur(2015)]{cates_motility-induced_2015}
M.~E. Cates and J.~Tailleur, 2015, \textbf{6}, 219--244\relax
\mciteBstWouldAddEndPuncttrue
\mciteSetBstMidEndSepPunct{\mcitedefaultmidpunct}
{\mcitedefaultendpunct}{\mcitedefaultseppunct}\relax
\EndOfBibitem
\bibitem[Bishop(1982)]{bishop1982collective}
M.~Bishop, \emph{Journal of Statistical Physics}, 1982, \textbf{29},
  623--629\relax
\mciteBstWouldAddEndPuncttrue
\mciteSetBstMidEndSepPunct{\mcitedefaultmidpunct}
{\mcitedefaultendpunct}{\mcitedefaultseppunct}\relax
\EndOfBibitem
\bibitem[Lepri \emph{et~al.}(2005)Lepri, Sandri, and Politi]{lepri2005one}
S.~Lepri, P.~Sandri and A.~Politi, \emph{The European Physical Journal
  B-Condensed Matter and Complex Systems}, 2005, \textbf{47}, 549--555\relax
\mciteBstWouldAddEndPuncttrue
\mciteSetBstMidEndSepPunct{\mcitedefaultmidpunct}
{\mcitedefaultendpunct}{\mcitedefaultseppunct}\relax
\EndOfBibitem
\bibitem[Lee-Dadswell \emph{et~al.}(2017)Lee-Dadswell, Barrett, and
  Power]{lee2017cluster}
G.~Lee-Dadswell, N.~Barrett and M.~Power, \emph{Physical Review E}, 2017,
  \textbf{96}, 032144\relax
\mciteBstWouldAddEndPuncttrue
\mciteSetBstMidEndSepPunct{\mcitedefaultmidpunct}
{\mcitedefaultendpunct}{\mcitedefaultseppunct}\relax
\EndOfBibitem
\bibitem[Srinivas and Bagchi(2000)]{Bagchi}
G.~Srinivas and B.~Bagchi, \emph{J.\ Chem. Phys.}, 2000, \textbf{112},
  7557\relax
\mciteBstWouldAddEndPuncttrue
\mciteSetBstMidEndSepPunct{\mcitedefaultmidpunct}
{\mcitedefaultendpunct}{\mcitedefaultseppunct}\relax
\EndOfBibitem
\bibitem[Lizana and Ambj{\"o}rnsson(2009)]{lizana2009diffusion}
L.~Lizana and T.~Ambj{\"o}rnsson, \emph{Physical Review E}, 2009, \textbf{80},
  051103\relax
\mciteBstWouldAddEndPuncttrue
\mciteSetBstMidEndSepPunct{\mcitedefaultmidpunct}
{\mcitedefaultendpunct}{\mcitedefaultseppunct}\relax
\EndOfBibitem
\bibitem[Campos and Oseguera(2019)]{campos}
C.~V. Campos and F.~A. Oseguera, \emph{arXiv:1912.01282}, 2019\relax
\mciteBstWouldAddEndPuncttrue
\mciteSetBstMidEndSepPunct{\mcitedefaultmidpunct}
{\mcitedefaultendpunct}{\mcitedefaultseppunct}\relax
\EndOfBibitem
\bibitem[Caprini \emph{et~al.}(2019)Caprini, Hern{\'a}ndez-Garc{\'i}a,
  L{\'o}pez, and Marconi]{Caprini}
L.~Caprini, E.~Hern{\'a}ndez-Garc{\'i}a, C.~L{\'o}pez and U.~M.~B. Marconi,
  \emph{Scientific Reports}, 2019, \textbf{9}, 16687\relax
\mciteBstWouldAddEndPuncttrue
\mciteSetBstMidEndSepPunct{\mcitedefaultmidpunct}
{\mcitedefaultendpunct}{\mcitedefaultseppunct}\relax
\EndOfBibitem
\bibitem[Quastel and Spohn(2015)]{quastel2015one}
J.~Quastel and H.~Spohn, \emph{Journal of Statistical Physics}, 2015,
  \textbf{160}, 965--984\relax
\mciteBstWouldAddEndPuncttrue
\mciteSetBstMidEndSepPunct{\mcitedefaultmidpunct}
{\mcitedefaultendpunct}{\mcitedefaultseppunct}\relax
\EndOfBibitem
\end{mcitethebibliography}
\bibliographystyle{rsc} %the RSC's .bst file

\end{document}